\documentclass[preprints,article,accept,moreauthors]{Definitions/mdpi}
\firstpage{1} 
\pubvolume{1}
\issuenum{1}
\articlenumber{0}
\pubyear{2022}
\copyrightyear{2022}
\datereceived{} 
\dateaccepted{} 
\datepublished{} 
\hreflink{https://doi.org/} 

\newcommand\fdg{\hbox{$.\!\!^\circ$}}
\newcommand\degr{\hbox{$^\circ$}}
\newcommand\sun{\hbox{$\odot$}}
\defcitealias{2011ApJS..197...21H}{CGS}
\defcitealias{RC3}{RC3}
\usepackage{dashrule}
\definecolor{MATLAB_blue}{rgb}{0,0.4470,0.7410}
\definecolor{MATLAB_red}{rgb}{0.8500,0.3250,0.0980}
\definecolor{MATLAB_orange}{rgb}{0.9290,0.6940,0.1250}
\definecolor{MATLAB_purple}{rgb}{0.4940,0.1840,0.5560}
\definecolor{MATLAB_green}{rgb}{0.4660,0.6740,0.1880}
\definecolor{MATLAB_cyan}{rgb}{0.3010,0.7450,0.9330}
\definecolor{MATLAB_maroon}{rgb}{0.6350,0.0780,0.1840}
\definecolor{LimeGreen}{rgb}{0.1961,0.8039,0.1961}
\usepackage{cleveref}
\usepackage{tabularx}
\Title{Probing the Low-mass End of the Black Hole Mass Function via a Study of Faint Local Spiral Galaxies}

\TitleCitation{Probing the Low-mass End of the Black Hole Mass Function via a Study of Faint Local Spiral Galaxies}

\Author{
Michael S.\ Fusco $^{1,2,*,\dagger}$\orcidA{},
Benjamin L.\ Davis $^{3,4}$\orcidB{},
Julia Kennefick $^{1,2}$\orcidC{},
Daniel Kennefick $^{1,2}$\orcidD{},
and Marc S.\ Seigar $^{5}$\orcidE{}
}

\AuthorNames{Michael S.\ Fusco, Benjamin L.\ Davis, Julia Kennefick, Daniel Kennefick, and Marc S.\ Seigar}

\AuthorCitation{Fusco, M.S.; Davis, B.L.; Kennefick, J.; Kennefick, D.; Seigar, M.S.}

\address{%
$^{1}$ \quad Arkansas Center for Space and Planetary Sciences, Fayetteville, AR 72701, USA\\
$^{2}$ \quad Department of Physics, University of Arkansas, Fayetteville, AR 72701, USA\\
$^{3}$ \quad Centre for Astrophysics and Supercomputing, Swinburne University of Technology, Hawthorn, VIC 3122, Australia\\
$^{4}$ \quad Center for Astro, Particle, and Planetary Physics (CAP$^3$), New York University Abu Dhabi, Abu Dhabi 129188, United Arab Emirates\\
$^{5}$ \quad Department of Physics and Astronomy, University of Toledo, Toledo, OH 43606, USA
}

\corres{Correspondence: michael.s.fusco@uconn.edu}

\firstnote{Current address: University of Connecticut Graduate School of Business, Storrs, CT 06269, USA.} 

\abstract{
We present an analysis of the pitch angle distribution function (PADF) for nearby galaxies and its resulting black hole mass function (BHMF) via the well-known relationship between pitch angle and black hole mass.
Our sample consists of a subset of 74 spiral galaxies from the Carnegie-Irvine Galaxy Survey with absolute $B$-band magnitude $\mathfrak{M}_{B}>-19.12$\,mag and luminosity distance $D_{\mathrm{L}} \leq 25.4$\,Mpc, which is an extension of a complementary set of 140 more luminous ($\mathfrak{M}_{B}\leq-19.12$\,mag) late-type galaxies.
We find the PADFs of the two samples are, somewhat surprisingly, not strongly dissimilar; a result that may hold important implications for spiral formation theories.
Our data show a distinct bimodal population manifest in the pitch angles of the Sa--Sc types and separately the Scd--Sm types, with Sa--Sc types having tighter spiral arms on average.
Importantly, we uncover a distinct bifurcation of the BHMF, such that the Sa--Sc galaxies typically host so-called ``supermassive'' black holes ($M_{\bullet}\gtrsim10^6\,\mathrm{M_{\sun}}$), whereas Scd--Sm galaxies accordingly harbor black holes that are ``less-than-supermassive'' ($M_{\bullet}\lesssim10^6\,\mathrm{M_{\sun}}$).
It is amongst this latter population of galaxies where we expect fruitful bounties of elusive intermediate-mass black holes (IMBHs), through which a better understanding will help form more precise benchmarks for future generations of gravitational wave detectors.
}


\begin{document}

\section{Introduction}\label{sec:intro}

     Supermassive black holes (SMBHs) reside at the centers of most galaxies \citep{1998Natur.395A..14R,2002Natur.419..694S}.
     There is strong evidence supporting proposed relationships among the mass of the central SMBH and various properties and features of their host galaxies \citep[for a thorough review, see][and references therein]{2015PKAS...30..335G}.
     Given that black hole masses correspond to morphological features of their host galaxies, a comprehensive census can be undertaken to characterise the distribution of SMBH masses, a black hole mass function (BHMF) in the local Universe.
     Extension of the BHMF to higher redshift galaxies would advance studies of galactic evolution through the time dependency of SMBH masses.
     The local census offers a snapshot of the current distribution of black hole masses in the present Universe while a higher redshift study would allow a mapping of the history of the BHMF through cosmic time.
     Incorporation of galaxies at greater redshifts will inform changes in the BHMF through processes such as mergers, accretion, and secular evolution.
	
	At present, the high-mass end of the BHMF has been probed through observations of the most active galaxies \citep{2009ApJ...690..537D}.
	In the local Universe, a majority of large black holes reside in elliptical or lenticular galaxies.
	Their black hole masses ($M_{\bullet}$) are estimated primarily through a correlation between their central stellar velocity dispersion, $\sigma$, and the mass of the SMBH \citep{2000ApJ...539L...9F,2000ApJ...539L..13G,Nicola:2019,Sahu:2019}.
	In general, the BHMF relies on applying scaling relations to luminosity/velocity functions, and has been utilized by many authors for populations of late-type and early-type galaxies \citep{1999MNRAS.307..637S,2002RvMA...15...57R,2004MNRAS.354.1020S,2007ApJ...663...53T,2009NewAR..53...57S}.
	
	For the low-mass end of the BHMF, there is still no clear consensus as to the behavior of the distribution \citep{2004MNRAS.351..169M,2004MNRAS.354.1020S,2009MNRAS.400.1451V}.
	A secure estimate of the low-mass end of the BHMF provides constraints on the degree of time evolution in the Eddington ratio distributions \citep[\textit{e.g.},][]{2009NewAR..53...57S}, as well as on the population of SMBH seeds \citep[\textit{e.g.},][]{2011arXiv1105.4902N}.
	To probe the evolution of the BHMF over the lifetime of the Universe, a more complete measurement of the distribution locally must be attained and modified to include galaxies with lower black hole masses.
	Therefore, our exploration continues a census of black holes in late-type galaxies, concentrating here on galaxies with lower absolute $B$-band luminosities.
	As we will see, these galaxies represent the low-mass tail of galactic black hole masses.
	
	The $M_{\bullet}$--$\sigma$ relation has been extensively adopted to estimate the black hole masses in early-type galaxies in order to establish a BHMF for those morphologies \citep{1998MNRAS.297..817F,2007MNRAS.378..198G}.
	The S{\'e}rsic index \citep{Sersic:1963,Sersic:1968,Graham:2005} is another widely-used black hole mass estimation method for early-type galaxies and the bulges of late-type galaxies \citep{Graham:2007,2013MNRAS.434..387S,Savorgnan:2016,Sahu:2020}.
	Spiral galaxies are structurally complex and require extensive multi-component decompositions before attempting to calculate accurate S{\'e}rsic indice values \citep{Stone:2021}.
	Moreover, the level scatter intrinsic to these methods is higher when used to estimate SMBH masses for late-type galaxies \citep{Sahu:2020}.
		
\begin{figure}
	 \includegraphics[clip=true, trim= 8mm 4mm 1mm 1mm, width=\textwidth]{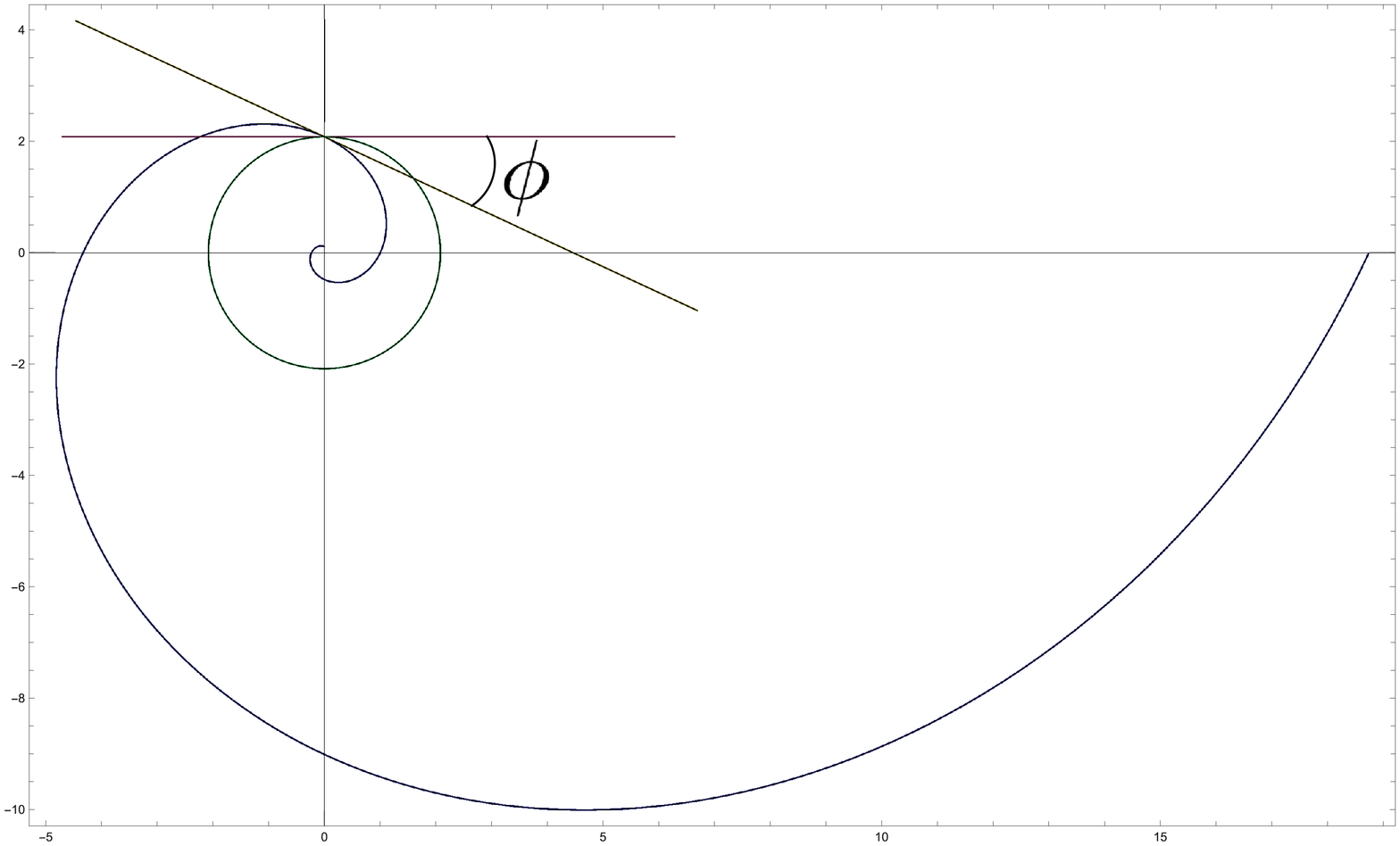}
	 \caption{
	 This figure illustrates the geometric definition of the pitch angle, $\phi$, of a logarithmic spiral.
	 Specifically, $\phi$ is the angle between the line drawn tangent to the circle and the other line drawn tangent to the logarithmic spiral, as depicted.
	 In this case, $\phi=-25\degr$, where the negative sign indicates, by our convention, that the spiral winds outward in a counterclockwise direction.
	 }
	 \label{fig:pitch}
	\end{figure}

	Here, we use a correlation between the SMBH mass and the logarithmic spiral pitch angle ($\phi$) of a galaxy, as proposed in \citet{Seigar:2008}, and developed further in \citet{2013ApJ...769..132B} and \citet{Davis:2017}.\endnote{
	We do offer a word of caution regarding all black hole mass scaling relations, including the $M_{\bullet}$--$\phi$ relation, in regards to estimating low-mass black holes.
	The most trustworthy scaling relations are built by direct, dynamical measurements of black holes.
	However, resolution requirements to directly measure black hole masses create an observational bias where only the nearest and biggest black holes have been measured.
	\citet{Sahu:2019} compiled the largest-to-date sample of galaxies with directly-measured black holes, which adds up to only 145 galaxies.
	Furthermore, if we consider only late-type galaxies, the lowest directly-measured black hole mass is found in NGC~4395, with measurements of $4.0_{-1.0}^{+2.7}\times10^5\,\mathrm{M_{\sun}}$ via gas dynamical modeling \citep{Brok:2015} and $2.5_{-0.8}^{+1.0}\times10^5\,\mathrm{M_{\sun}}$ via gas kinematics \citep{Brum:2019}.
	Thus, any black hole mass estimate produced by the $M_{\bullet}$--$\phi$ relation, or any other black hole mass scaling relation for late-type galaxies, below the black hole mass of NGC~4395 is by definition an extrapolation.
	Until such a time as more low-mass black holes are directly measured, mass predictions below $\approx$$10^6\,\mathrm{M_{\sun}}$ are uncertain, and below $10^5\,\mathrm{M_{\sun}}$ are speculation.
	}
	See Figure~\ref{fig:pitch} for a geometric definition of $\phi$, and for a mathematical definition, see \citet[][\S2]{Davis:2017}.
	According to the $M_{\bullet}$--$\phi$ relation, galaxies with larger SMBH masses will have more tightly wound spiral arms with lower pitch angles, whereas galaxies with lower SMBH masses will have looser spirals with higher pitch angles.
	This $M_{\bullet}$--$\phi$ relation has a few advantages over past techniques for measurement of SMBH masses in late-type galaxies.
	The $M_{\bullet}$--$\phi$ relation has a tighter correlation than other methods when applied to late-type galaxies.
	More importantly, measuring the pitch angle of galaxy spiral arms only requires sufficiently resolved images of the galaxy rather than observationally-intensive spectra.
	
	It may be, then, the best method for extending SMBH mass estimation in late-type galaxies to higher redshifts is through the $M_{\bullet}$--$\phi$ relation rather than other current methods.\endnote{
	As pointed out in \citet{Davis:2018}, the total stellar mass of a galaxy ($M_{\mathrm{\star,tot}}$) produces an $M_{\bullet}$--$M_\mathrm{\star,tot}$ relation with a level of scatter that is on par with the spheroid (bulge) stellar mass relation, $M_{\bullet}$--$M_\mathrm{\star,sph}$.
	Moreover, $M_\mathrm{\star,tot}$ is considerably easier to measure than $M_\mathrm{\star,sph}$ because it is only considered with measuring the total light of a galaxy rather than carefully decomposing the light profile a galaxy to yield accurate bulge masses.
	As such, the $M_{\bullet}$--$M_\mathrm{\star,tot}$ relation is a great choice for late-type galaxies because it works even for bulgeless galaxies and is useful at higher redshifts where resolving the components of a galaxy becomes increasingly difficult.
	However, in their \S2.3, \citet{Davis:2018} remark that the $M_{\bullet}$--$M_\mathrm{\star,tot}$ relation is only as accurate as the mass-to-light ratios used to convert the total light of a galaxy into an equivalent stellar mass.
	Plagued with a myriad of complexities \citep[\textit{e.g.},][]{Graham:2022,Sahu:2022}, the accurate estimation of mass-to-light ratios (and their heterogenous adoption, akin to the ``little $h$'' of cosmology) is perhaps one of the most pernicious problems in extragalactic astronomy.
	}
	As a result, the $M_{\bullet}$--$\phi$ relation is a useful method both for measuring the SMBH mass of spiral galaxies locally and at higher redshift, with the caveat that spiral structure be sufficiently resolved in imaging.
	For example, \citet{Shields:2015} find that pitch angles are reliably measured out to $z=1.2$ for spiral galaxies in he Great Observatories Origins Deep Survey \citep{Giavalisco:2004}.
	In recent years, pitch angles have been measured for spiral galaxies spotted at incredible redshifts: $z=2.54$ \citep{Yuan:2017} and $z=4.41$ \citep{Tsukui:2021}.
	These findings have opened up a new world of exploration for spiral galaxies.
	Just a few years ago, spiral galaxies were thought not to exist beyond $z=2$, and the spirals that were found from that epoch were thought not to be \emph{true} spirals, just conglomerations of clumpy gas clouds \citep[\textit{e.g.},][]{Elmegreen:2014}.
    
    Pitch angle is also correlated with the mass of the central bulge region of spiral galaxies \citep{2019ApJ...873...85D}.
    Modal density wave theory \citep{Lindblad:1963,1964ApJ...140..646L} describes spiral structure in galaxies as a standing wave pattern of density waves.
    These density waves are generated by inner and outer Lindblad resonance orbits in a galaxy \citep{Lindblad:1948,Lindblad:1960,Lindblad:1963,1964ApJ...140..646L}.
    They propagate through the disk, amplified or damped by interactions with galactic features.
    The pitch angle of the spiral arms is analogous to the wavelength of a standing wave on a vibrating string.
    Specifically, the absolute value of pitch angle is dependent on the ratio of the mass density in the disk to the mass in the central region \citep{Lin:1966}.
    Thus, the SMBH mass behaves like the tension on a string, while the disk itself acts as the medium of the string.
    
    This formulation of modal density wave theory has been applied effectively to the behavior of spirals in the rings of Saturn \citep{1981Natur.292..703C}.
    Spiral galaxies, though more structurally complex than the high central mass, low disk mass ideal case, nevertheless have a configuration of a centrally concentrated mass (SMBH and bulge) with a thin disk of orbiting material.
    Since the mass of the galactic bulge correlates well with the SMBH mass \citep[\textit{e.g.},][]{2019ApJ...873...85D,Sahu:2019a}, it follows that application of modal density wave theory to indirectly measure SMBH masses is a viable method.
    Unlike the end-member case of Saturn, where the disk mass is much less than the central mass, spiral galaxies have a comparatively flat mass distribution as evidenced in their rotation curves.\endnote{Volute structures induced by density waves are also present in protoplanetary and circumplanetary disks \citep{2016Sci...353.1519P, 2018MNRAS.480.4327X, 2016ApJ...833..126B}, and \citet{Yu:2019b} show a tight relation between $|\phi|$ and protoplanetary disk mass.}
    As the mass ratio of central mass to disk mass decreases, the logarithmic spiral-arm pitch angle increases.
    While Saturn's ring spirals have a pitch angle $\ll$$1\degr$, most spiral galaxies are closer to $\approx$$20\degr$.
    It has been shown in \citet{Davis:2015}, there is a planar relationship among the quantities of bulge stellar mass, $\phi$, and the mass of neutral hydrogen (proxy for disk density) in a galaxy, in accordance with density wave theory.
   
    A first census of the BHMF in spiral galaxies has been conducted by \citet{2014ApJ...789..124D}, in which a volume-limited sample of spiral galaxies was selected from the Carnegie-Irvine Galaxy Survey \citep[][hereafter \citetalias{2011ApJS..197...21H}]{2011ApJS..197...21H} sample of nearby galaxies with two bounds, luminosity-distance and absolute $B$-band magnitude.
    The reasoning for these bounds was for sample completeness, as fainter galaxies might not be identified beyond a specified distance.
    Here, we consider these fainter galaxies in the larger sample, accounting for missing galaxies by weighted correction.
    
    We define our sample of galaxies in \S\ref{sec:Data}.
    In \S\ref{sec:Analysis}, we present our pitch angle measurements and properties of the sample.
    We present the pitch angle distribution function (PADF) and BHMF in \S\ref{sec:PADF} and \S\ref{sec:BHMF}, respectively.
    Finally, we provide a discussion of our results (\S\ref{sec:discussion}) and conclusions (\S\ref{sec:Conclusions}).
    Unless otherwise stated, all uncertainties are quoted at $1\,\sigma \approx 68.3\%$ confidence intervals.
    Except when mentioning individual galaxies, we casually refer to Hubble type categories without reference to bar morphologies, \textit{i.e.}, ``Sd'' (or just ``d'') is used to generally include SAd, SABd, and SBd morphologies.

\section{Methodology \& Data}\label{sec:Data}
	\subsection{Sample Selection}
	
	Following \citet{2014ApJ...789..124D}, we make use of Southern Hemisphere galaxies based on the \citetalias{2011ApJS..197...21H}.
    The \citetalias{2011ApJS..197...21H} was a project to observe 605 bright southern galaxies using the Las Campanas Observatory in each of the Johnson \textit{B, V, R}, and $I$ filters.\endnote{
    We note that the \citetalias{2011ApJS..197...21H} sample includes the Large Magellanic Cloud (ESO~056-G115), but not the irregular Small Magellanic Cloud (NGC~292).}
    For this research, we utilize optical ($B$-band) imaging of a subsample of these galaxies from the NASA/IPAC Extragalactic Database (NED).\endnote{\url{https://ned.ipac.caltech.edu/}}
	Pursuit of a BHMF for the local Universe is undertaken through the measurement of pitch angles to estimate black hole masses.
	
	The formerly defined \citet{2014ApJ...789..124D} sample provided an initial construction of the BHMF of late-type galaxies, wherein the pitch angles of a volume-limited set of local spiral galaxies were measured to calculate the BHMF.
	Their dataset excluded galaxies below a specified luminosity; beyond the distance limit chosen, some fraction of these galaxies were beyond the detection limit of the survey.
	Consequently, their dataset is complete (no undetected galaxies) within its distance and luminosity bounds.
	Yet, these limitations necessarily exclude the lowest luminosity galaxies, which may comprise the extreme low-mass end of the BHMF.
	We measure pitch angles for all galaxies detected within the luminosity-distance limit, accounting for those too faint to be identified.
	From hereon, we will refer to their sample as the \emph{bright} sample \citep{2014ApJ...789..124D} and our sample as the \emph{faint} sample.

\begin{figure}[h]
	    \includegraphics[clip=true, trim= 2mm 0mm 4mm 3.8mm, width=\textwidth]{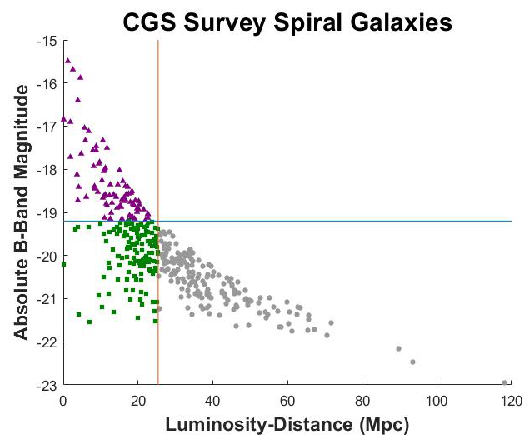}
	    \caption{
	    Luminosity distance \textit{vs.}\ absolute $B$-band magnitude for the Carnegie-Irvine Galaxy Survey \citepalias{2011ApJS..197...21H} galaxies.
	    The \citet{2014ApJ...789..124D} sample (green squares, \textcolor{LimeGreen}{$\blacksquare$}) and our faint galaxy sample (purple triangles, \textcolor{violet}{$\blacktriangle$}) are illustrated.
	    The vertical (\textcolor{orange}{{\hdashrule[0.35ex]{9mm}{0.4mm}{}}}) line is positioned at $D_\mathrm{L} = 25.4$\,Mpc (gray circles, \textcolor{lightgray}{$\bullet$} right of this line represent galaxies that are too distant and thus, not included in either sample) and the horizontal (\textcolor{cyan}{{\hdashrule[0.35ex]{9mm}{0.4mm}{}}}) line is positioned at $\mathfrak{M}_{B}=-19.12$\,mag.
	    }
	\label{fig:selection}
	\end{figure}

	
	Our sample of galaxies draws 74 spiral galaxies from the \citetalias{2011ApJS..197...21H}, with absolute $B$-band magnitude $\mathfrak{M}_{B}>-19.12$\,mag\endnote{
	This is equivalent to a luminosity of $\log(L/{\rm L}_{\sun})<9.78$, assuming the absolute magnitude of the Sun in the Johnson $B$-band is 5.44\,mag in the \textit{vegamag} system \citep{Willmer:2018}.} and limiting luminosity (redshift-independent) distance $D_\mathrm{L} \leq 25.4$\,Mpc ($z\leq0.00572$).
	For a derivation of these sample boundaries, see \citet[][\S3]{2014ApJ...789..124D}.
	Figure~\ref{fig:selection} shows the full sample consisting of 140 spiral galaxies (volume- and magnitude-limited) from \citet{2014ApJ...789..124D}, plus our 74 faint galaxies in this study (still volume-limited, but not magnitude-limited).
	To remain consistent with \citet{2014ApJ...789..124D}, we adopted the same cosmographic parameters from \citet{2014A&A...566A..54P}, with $\omega_{\rm b} = 0.022161$, $\Omega_{\rm M} = 0.3071$, $\Omega_{\Lambda} = 0.6914$, and $h_{67.77}  = h/0.6777 = H_{0}/(67.77$ km s$^{-1}$\,Mpc$^{-1}) \equiv 1$.
	Defines our sample with the same comoving volume, $V_{\rm C} = 3.37$ $\times$ $10^4$\,$h_{67.77}^{-3}$\,Mpc$^3$ and lookback time, $t_{\rm L} \leq 82.1$\,$h_{67.77}^{-1}$\,Myr.
	
	Of the 74 faint galaxies, galactic spiral-arm pitch angles were measured for 55 galaxies.
	Of the 19 galaxies which we were unable to measure, the ``Hubble'' type \citep{Jeans:1919,Jeans:1928,Lundmark:1925,Hubble:1926,Hubble:1926b,Hubble:1927,Hubble:1936} of the galaxy being examined greatly effected the measurement success rate.
	Measurement success did not depend strongly on the $\mathfrak{M}_{B}$ of the galaxies examined, as measurement failures occurred at all magnitudes.
	However, no galaxies fainter than $\mathfrak{M}_{B}=-17$\,mag were successfully measured.
	The characteristics of our measured galaxies are included in Table~\ref{table}.
	
	
\begin{table}
\caption{The subsample of 74 spiral galaxies (55 measured $+$ 19 with unmeasurable pitch angles) from the \citetalias{2011ApJS..197...21H} sample.}
	\label{table}
	\begin{adjustwidth}{-4cm}{0cm}
		\begin{tabularx}{\fulllength}{llrrrCCrC}
			\toprule
			\multicolumn{1}{c}{Galaxy}      & \multicolumn{1}{c}{Hubble Type}  & \multicolumn{1}{c}{$B_{\mathrm{T}}$}  & \multicolumn{1}{c}{$D_{\mathrm{L}}$} & \multicolumn{1}{c}{$A_{B}$}  & $\mathfrak{M}_{B}$   & $\log(L/{\rm L}_{\sun})$ & \multicolumn{1}{c}{$\phi$} & $\log(M_{\bullet}/{\rm M}_{\sun})$ \\
			 &  & \multicolumn{1}{c}{[mag]} & \multicolumn{1}{c}{[Mpc]} & \multicolumn{1}{c}{[mag]} & \multicolumn{1}{c}{[mag]} &  & \multicolumn{1}{c}{[deg]} &  \\
			\multicolumn{1}{c}{(1)} & \multicolumn{1}{c}{(2)} & \multicolumn{1}{c}{(3)} & \multicolumn{1}{c}{(4)} & \multicolumn{1}{c}{(5)} & \multicolumn{1}{c}{(6)} & \multicolumn{1}{c}{(7)} & \multicolumn{1}{c}{(8)} & \multicolumn{1}{c}{(9)}  \\
			\midrule
			ESO 056-G115	& SBm	& 1.92 	& 0.050	& 0.272	& $-16.85$ & 8.87               & \textemdash					& \textemdash	       \\
			ESO 137-G018	& Sc		& 12.40 	& 5.867	& 0.882	& $-17.32$ & 9.06               & \textemdash					& \textemdash	       \\
			ESO 219-G021	& SABc	& 12.83 	& 14.220	& 0.707	& $-18.64$ & 9.59               & $-19.93\pm\phantom{1}2.36$	& $6.97\pm0.47$       \\
			ESO 265-G007	& SBc	& 12.54 	& 13.400  	& 0.498	& $-18.59$ & 9.57               & $16.41\pm\phantom{1}2.12$    	& $7.19\pm0.46$       \\
			ESO 270-G017	& SBm	& 12.13 	& 5.676	& 0.401	& $-17.04$ & 8.95               & \textemdash					& \textemdash	       \\
			ESO 271-G010	& Sc		& 12.90 	& 18.889 	& 0.355	& $-18.84$ & 9.66               & $29.54\pm\phantom{1}9.45$    	& $6.38\pm0.76$       \\
			ESO 274-G001	& Scd	& 11.73 	& 4.874	& 0.930	& $-17.64$ & 9.19               & \textemdash					& \textemdash	       \\
			ESO 358-G063	& Sc		& 12.57 	& 19.673	& 0.023	& $-18.92$ & 9.70               & \textemdash					& \textemdash	       \\
			ESO 362-G011	& Sbc	& 12.77 	& 20.217	& 0.174	& $-18.93$ & 9.70               & \textemdash					& \textemdash	       \\
			ESO 373-G008	& Sc		& 12.71 	& 11.100	& 0.484	& $-18.00$ & 9.33               & \textemdash					& \textemdash	       \\
			ESO 383-G087	& Sd		& 11.70 	& 2.675    	& 0.257 	& $-15.69$ & 8.41               & $9.59\pm\phantom{1}0.76$   	& $7.62\pm0.42$       \\
			ESO 384-G002	& SBd	& 12.88 	& 17.450  	& 0.232 	& $-18.56$ & 9.55               & $-11.89\pm\phantom{1}1.45$    	& $7.47\pm0.44$       \\
			ESO 479-G004	& SBd  	& 12.79 	& 19.400  	& 0.066 	& $-18.72$ & 9.62               & $28.12\pm\phantom{1}3.40$    	& $6.47\pm0.53$       \\
			IC 1954      	& Sb      	& 12.10 	& 15.287	& 0.059 	& $-18.88$ & 9.68               & $18.32\pm\phantom{1}2.70$    	& $7.07\pm0.47$       \\
			IC 1993      	& SABb    & 12.52 	& 14.700	& 0.036 	& $-18.35$ & 9.47               & $-20.46\pm\phantom{1}2.48$    	& $6.94\pm0.48$       \\
			IC 2000      	& SBc      	& 12.82 	& 20.080  	& 0.036 	& $-18.73$ & 9.62               & $-24.14\pm\phantom{1}1.93$    	& $6.71\pm0.48$       \\
			IC 2056      	& Sbc     	& 12.80 	& 20.500	& 0.067 	& $-18.83$ & 9.66               & $16.34\pm\phantom{1}5.06$    	& $7.20\pm0.54$       \\
			IC 2627     	& SABb    & 12.68 	& 18.480 	& 0.436 	& $-19.09$ & 9.77               & $-33.18\pm23.55$   			& $6.15\pm1.55$       \\
			IC 4710		& Sm	& 12.50 	& 8.900	& 0.322	& $-17.57$ & 9.16               & \textemdash					& \textemdash	       \\
			IC 5052		& SBcd	& 11.79 	& 8.095	& 0.184	& $-17.94$ & 9.31               & \textemdash					& \textemdash	       \\
			IC 5201     	& Sc     	& 11.95 	& 14.400 	& 0.043 	& $-18.88$ & 9.68               & $34.92\pm23.11$   			& $6.04\pm1.52$       \\
			IC 5273     	& SBc    	& 12.74 	& 16.557 	& 0.045 	& $-18.40$ & 9.49               & $21.37\pm\phantom{1}6.88$    	& $6.89\pm0.62$       \\
			IC 5332     	& SABc  	& 11.23 	& 8.400 	& 0.060 	& $-18.45$ & 9.51               & $4.99\pm\phantom{1}0.39$    	& $7.90\pm0.42$       \\
			NGC 24       	& Sc       	& 12.14 	& 6.889  	& 0.071 	& $-17.12$ & 8.98               & $-25.40\pm23.11$   			& $6.64\pm1.51$       \\
			NGC 45       	& SABd 	& 11.39 	& 9.313 	& 0.075 	& $-18.53$ & 9.54               & $-27.98\pm\phantom{1}3.97$    	& $6.48\pm0.54$       \\
			NGC 55		& SBm	& 9.59 	& 1.942	& 0.048	& $-16.90$ & 8.89               & \textemdash					& \textemdash	       \\
			NGC 247      	& SABc 	& 9.72  	& 3.587	& 0.066 	& $-18.12$ & 9.38               & $42.90\pm\phantom{1}4.84$    	& $5.55\pm0.64$       \\
			NGC 300      	& Scd    	& 8.81  	& 1.972 	& 0.046 	& $-17.71$ & 9.22               & $33.69\pm\phantom{1}0.75$    	& $6.12\pm0.51$       \\
			NGC 625		& SBm	& 11.57 	& 3.807	& 0.060	& $-16.39$ & 8.69               & \textemdash					& \textemdash	       \\
			NGC 701      	& SBc    	& 12.84 	& 22.711	& 0.091 	& $-19.03$ & 9.74               & $17.15\pm\phantom{1}1.78$    	& $7.15\pm0.45$       \\
			NGC 779      	& SABb 	& 12.27 	& 17.678	& 0.097 	& $-19.06$ & 9.76               & $15.58\pm\phantom{1}5.52$    	& $7.24\pm0.55$       \\
			NGC 1022    	& SBa    	& 12.32 	& 18.500	& 0.093 	& $-19.11$ & 9.77               & $22.44\pm12.10$   			& $6.82\pm0.88$ \\
			NGC 1042     	& SABc 	& 12.11 	& 9.431	& 0.104 	& $-17.87$ & 9.28               & $20.71\pm\phantom{1}6.55$    	& $6.93\pm0.61$ \\
			NGC 1179     	& Sc       	& 12.83 	& 18.171	& 0.087 	& $-18.55$ & 9.55               & $-19.18\pm\phantom{1}5.32$    	& $7.02\pm0.56$ \\
			NGC 1249     	& SBc      	& 12.78 	& 15.854	& 0.060 	& $-18.28$ & 9.44               & $-17.13\pm\phantom{1}5.93$    	& $7.15\pm0.57$ \\
			NGC 1292     	& Sc       	& 12.79 	& 21.475	& 0.062 	& $-18.93$ & 9.70               & $-14.87\pm\phantom{1}3.37$    	& $7.29\pm0.48$ \\
			NGC 1313     	& SBcd 	& 9.66  	& 3.951	& 0.396 	& $-18.72$ & 9.62               & $-41.63\pm\phantom{1}1.08$    	& $5.63\pm0.56$ \\
			NGC 1337     	& SBc      	& 12.53 	& 16.075	& 0.245 	& $-18.75$ & 9.63               & $-17.57\pm\phantom{1}5.26$   	& $7.12\pm0.55$ \\
			NGC 1436     	& Sb       	& 12.86 	& 18.360	& 0.039 	& $-18.50$ & 9.53               & $-24.58\pm\phantom{1}4.48$    	& $6.69\pm0.54$ \\
			NGC 1487     	& Scd     	& 12.33 	& 8.900 	& 0.043 	& $-17.46$ & 9.12               & $-69.44\pm\phantom{1}2.21$    	& $3.90\pm0.76$ \\
			NGC 1493     	& SBc      	& 11.72 	& 11.300	& 0.037 	& $-18.58$ & 9.56               & $14.21\pm\phantom{1}3.67$    	& $7.33\pm0.49$ \\
			NGC 1494     	& Scd     	& 12.10 	& 15.217	& 0.022 	& $-18.83$ & 9.66               & $22.01\pm\phantom{1}2.05$    	& $6.85\pm0.47$ \\
			NGC 1507	& SBd	& 12.86 	& 11.171	& 0.595	& $-17.98$ & 9.32               & \textemdash					& \textemdash \\
			NGC 1518     	& SBd      	& 12.26 	& 9.596 	& 0.174 	& $-17.82$ & 9.26               & $-46.01\pm\phantom{1}9.83$    	& $5.36\pm0.84$ \\
			NGC 1637     	& Sc       	& 11.63 	& 10.703	& 0.146 	& $-18.66$ & 9.60               & $13.76\pm\phantom{1}1.68$    	& $7.36\pm0.44$ \\
			NGC 1688     	& SBcd    	& 12.20 	& 15.450 	& 0.125 	& $-18.87$ & 9.68               & $-38.66\pm\phantom{1}0.95$    	& $5.81\pm0.54$ \\
			NGC 1744     	& SBc      	& 11.71 	& 10.853	& 0.148 	& $-18.62$ & 9.58               & $10.61\pm\phantom{1}3.75$    	& $7.55\pm0.48$ \\
			NGC 1796     	& SBc      	& 12.88 	& 10.600	& 0.088 	& $-17.33$ & 9.07               & $15.02\pm\phantom{1}2.92$    	& $7.28\pm0.47$ \\
			NGC 1892	& Sc		& 12.83 	& 16.900	& 0.272	& $-18.58$ & 9.56               & \textemdash					& \textemdash \\
			NGC 2082     	& SBb      	& 12.73 	& 17.978	& 0.210 	& $-18.75$ & 9.63               & $22.26\pm\phantom{1}2.41$    	& $6.83\pm0.48$ \\
			NGC 2090     	& Sbc     	& 11.66 	& 12.758	& 0.144 	& $-19.01$ & 9.73               & $8.29\pm\phantom{1}2.37$    	& $7.70\pm0.44$ \\
			\bottomrule
		\end{tabularx}
	\end{adjustwidth}
	\smallskip
\footnotesize
\emph{Continued on a subsequent page.}
\end{table}

\setcounter{table}{0}
\begin{table}[t]
\caption{\emph{Continued}}
	\begin{adjustwidth}{-4cm}{0cm}
		\begin{tabularx}{\fulllength}{llrrrCCrC}
			\toprule
			\multicolumn{1}{c}{Galaxy}      & \multicolumn{1}{c}{Hubble Type}  & \multicolumn{1}{c}{$B_{\mathrm{T}}$}  & \multicolumn{1}{c}{$D_{\mathrm{L}}$} & \multicolumn{1}{c}{$A_{B}$}  & $\mathfrak{M}_{B}$   & $\log(L/{\rm L}_{{\sun}})$ & \multicolumn{1}{c}{$\phi$} & $\log(M_{\bullet}/{\rm M}_{{\sun}})$ \\
			 &  & \multicolumn{1}{c}{[mag]} & \multicolumn{1}{c}{[Mpc]} & \multicolumn{1}{c}{[mag]} & \multicolumn{1}{c}{[mag]} &  & \multicolumn{1}{c}{[deg]} &  \\
			\multicolumn{1}{c}{(1)} & \multicolumn{1}{c}{(2)} & \multicolumn{1}{c}{(3)} & \multicolumn{1}{c}{(4)} & \multicolumn{1}{c}{(5)} & \multicolumn{1}{c}{(6)} & \multicolumn{1}{c}{(7)} & \multicolumn{1}{c}{(8)} & \multicolumn{1}{c}{(9)}  \\
			\midrule
			NGC 2188	& SBm	& 12.04 	& 7.900	& 0.118	& $-17.57$ & 9.16               & \textemdash					& \textemdash \\
			NGC 2427     	& SABd    & 12.30 	& 11.782 	& 0.778 	& $-18.83$ & 9.66               & $20.87\pm\phantom{1}3.62$    	& $6.92\pm0.51$ \\
			NGC 3109	& SBm	& 10.35 	& 1.324	& 0.242	& $-15.50$ & 8.34               & \textemdash					& \textemdash \\
			NGC 3513     	& SBc      	& 12.04 	& 13.153	& 0.230  	& $-18.79$ & 9.64               & $28.76\pm\phantom{1}0.56$    	& $6.43\pm0.49$ \\
			NGC 4592     	& Sd       	& 12.90  	& 11.630 	& 0.082 	& $-17.51$ & 9.14               & $40.30\pm17.93$   			& $5.71\pm1.24$ \\
			NGC 4632     	& Sc       	& 12.63 	& 18.518	& 0.087 	& $-18.79$ & 9.65               & $22.92\pm\phantom{1}3.74$    	& $6.79\pm0.52$ \\
			NGC 4700	& SBc	& 12.44 	& 15.62	& 0.170	& $-18.70$ & 9.61               & \textemdash					& \textemdash \\
			NGC 4951     	& SABc	& 12.58 	& 16.600	& 0.171 	& $-18.69$ & 9.61               & $18.51\pm\phantom{1}5.94$    	& $7.06\pm0.58$ \\
			NGC 5068     	& Sc       	& 10.64 	& 6.075 	& 0.369 	& $-18.65$ & 9.59               & $23.60\pm13.00$				& $6.75\pm0.93$ \\
			NGC 5134     	& SABb    & 12.46 	& 10.889 	& 0.328 	& $-18.05$ & 9.35               & $-36.28\pm\phantom{1}5.67$	& $5.96\pm0.63$ \\
			NGC 5264	& SBm	& 12.55 	& 4.467	& 0.187	& $-15.89$ & 8.49               & \textemdash					& \textemdash \\
			NGC 5556     	& Scd     	& 12.81 	& 18.750	& 0.254 	& $-18.81$ & 9.65               & $-14.09\pm\phantom{1}1.31$ 	& $7.34\pm0.44$ \\
			NGC 7090	& Sc		& 11.31 	& 8.399	& 0.083	& $-18.39$ & 9.49               & \textemdash					& \textemdash \\
			NGC 7361     	& Sc       	& 12.82 	& 17.200 	& 0.060	& $-18.42$ & 9.50               & $24.12\pm\phantom{1}7.22$  	& $6.71\pm0.65$ \\
			NGC 7412     	& SBb      	& 11.92 	& 12.485	& 0.042 	& $-18.60$ & 9.57               & $14.66\pm\phantom{1}2.23$   	& $7.30\pm0.45$ \\
			NGC 7421     	& Sbc     	& 12.78 	& 22.600	& 0.054 	& $-19.04$ & 9.75               & $20.88\pm\phantom{1}7.18$   	& $6.92\pm0.64$ \\
			NGC 7424     	& Sc       	& 11.54 	& 11.500 	& 0.039 	& $-18.80$ & 9.65               & $20.24\pm\phantom{1}6.05$   	& $6.96\pm0.59$ \\
			NGC 7456     	& Sc       	& 12.43	& 16.185 	& 0.039 	& $-18.65$ & 9.59               & $-18.48\pm\phantom{1}7.27$  	& $7.06\pm0.63$ \\
			NGC 7496     	& Sb       	& 12.90  	& 15.020	& 0.036 	& $-18.02$ & 9.34               & $-36.03\pm\phantom{1}0.84$  	& $5.98\pm0.53$ \\
			NGC 7513	& SBb	& 12.64 	& 19.85	& 0.148	& $-19.00$ & 9.73               & \textemdash		 			& \textemdash \\
			NGC 7713     	& Scd     	& 11.87 	& 10.272 	& 0.060  	& $-18.25$ & 9.43               & $-21.62\pm\phantom{1}2.78$  	& $6.87\pm0.49$ \\
			NGC 7793     	& Scd     	& 9.77  	& 4.171 	& 0.071	& $-18.40$ & 9.49               & $11.37\pm\phantom{1}1.21$   	& $7.51\pm0.43$ \\
			PGC\;\,3853     	& SABc    & 12.62 	& 12.600 	& 0.480  	& $-18.36$ & 9.48               & $23.17\pm\phantom{1}5.45$   	& $6.77\pm0.57$ \\
			\bottomrule
		\end{tabularx}
	\end{adjustwidth}
		\smallskip
\footnotesize
	\underline{Columns}:
	(1)~Galaxy Name.
	(2)~Hubble type morphological classification.
	(3)~$B$-band total magnitude.
	(4)~Luminosity distance (in Mpc).
	(5)~$B$-band extinction magnitude. 
	(6)~$B$-band absolute magnitude.
	(7)~Luminosity, assuming the absolute magnitude of the Sun in the Johnson $B$-band is 5.44\,mag in the \textit{vegamag} system \citep{Willmer:2018}.
	(8)~$B$-band pitch angle (in degrees), with $+\phi\rightarrow$``S''-wise winding and $-\phi\rightarrow$``Z''-wise winding.
	(9)~Black hole mass calculated from Equation~\ref{eqn:B+13}.
	\emph{Note}~--~The entries found in Columns~(1)--(5) are reproduced from the \citetalias{2011ApJS..197...21H} and derived from references therein.
\end{table}

Of course, the late-type-only BHMF by \citet{2014ApJ...789..124D} is not complete because it did not included early-type galaxies.
Therefore, \citet{Mutlu-Pakdil:2016} complemented the study by deriving a BHMF from all of the \citetalias{2011ApJS..197...21H} early-type galaxies included in the sample selection criteria of \citet{2014ApJ...789..124D}.
Thus, \citet{Mutlu-Pakdil:2016} added 68 early-type galaxies to the BHMF to generate a complete BHMF from ($68+140=$) 208 galaxies.
Our stated goal in this work is to probe the low-mass end of the BHMF. 
Inasmuch, we endeavor to further extend the late-type BHMF of \citet{2014ApJ...789..124D}, because low-mass central black holes are more likely to reside in low-mass spiral galaxies rather than elliptical and lenticular galaxies, which are typically more massive.
Similarly, a follow-up work to probe the low-mass end of the early-type BHMF from \citet{Mutlu-Pakdil:2016} could be conducted, but that is beyond the scope and immediate goals of this study.
		
	\subsection{Sample Comparison}
	
\begin{figure}
    \begin{adjustwidth}{-4cm}{0cm}
		\includegraphics[clip=true, trim= 3mm 1mm 4mm 9mm, width=9.23cm]{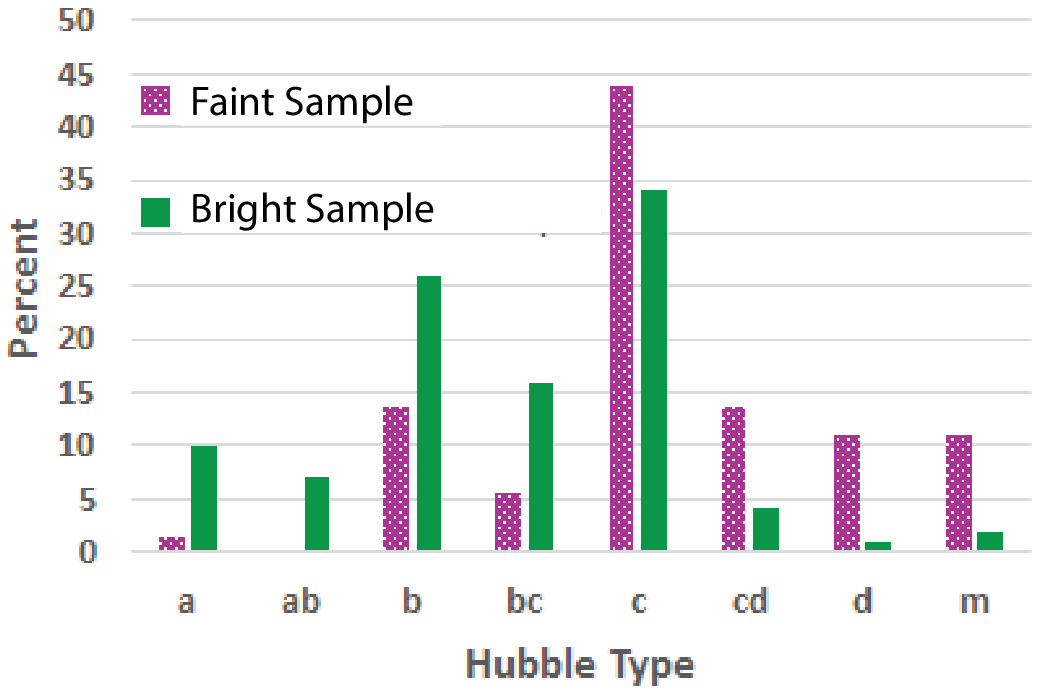}
		\includegraphics[clip=true, trim= 8mm 0mm 15mm 21mm, width=9.23cm]{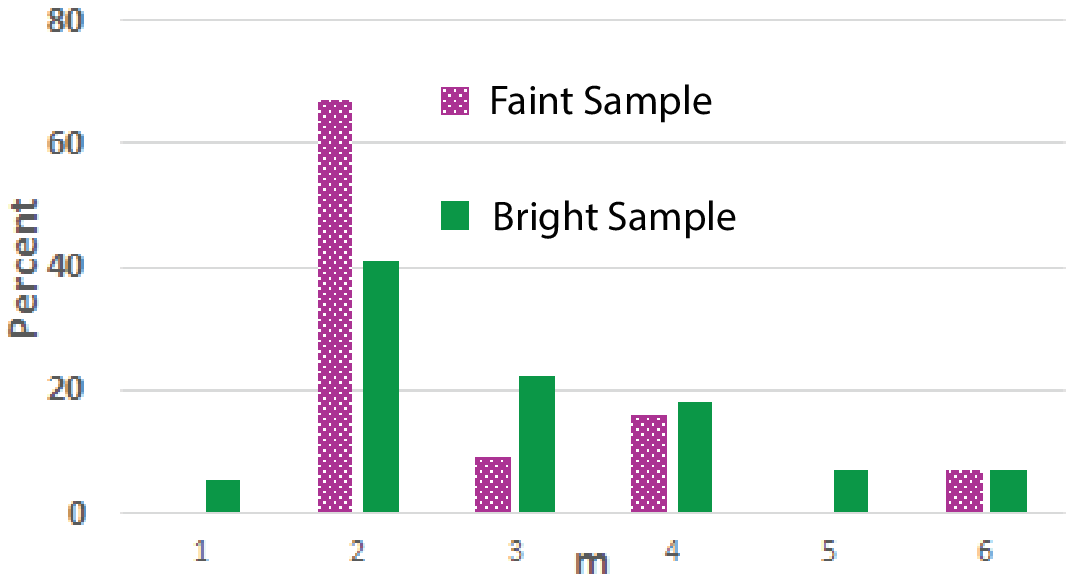}
		\end{adjustwidth}
		\caption{
		(\textbf{Left}) Histogram of Hubble types for all galaxies from each sample.
		The bright sample is comprised of spirals of medium arm winding (Sb--Sc), while the faint sample favors less bulge-dominated spirals with a high-$|\phi|$ sub-population (Sc--Sm).
		The average Hubble state for the bright sample is $T=3.8\pm1.6$, \textit{cf.} $T=5.4\pm1.7$ for the dim sample.
		Hence, a morphological difference is expressed between the two selections of galaxies.
		(\textbf{Right}) The histogram of the number of arms ($m$) for the two samples are similar, with two-armed ($m=2$) galaxies being most abundant in both.
		Notably, the faint sample is partially comprised of flocculent galaxies which often evade pitch angle measurement or arm number attribution, as opposed to the more accurately measured grand design spirals, which are common to the earlier \citet{2014ApJ...789..124D} sample.
		}
		\label{fig:types}
	\end{figure}

    Here, we consider the physical properties of the \citet{2014ApJ...789..124D} sample compared to our fainter set.
    Two such properties are the Hubble type classification and harmonic modality (number of arms) of the galaxies (see Figure~\ref{fig:types}).
    Differences in these characteristics are indicators of distinct morphological populations of spiral galaxies comprising the two samples.
    In comparing the two sets of galaxies, the faint sample contains many more Sc and Sd type spirals, implying less bulge-dominated and looser spiral winding in the galaxies, on average.
    According to the Tully-Fisher relation \citep{1977A&A....54..661T} the luminosity of a spiral galaxy relates to its rotational velocity\endnote{Furthermore, \citet{Kennicutt:1981} showed that $|\phi|$ is correlated with the rotational velocity of a spiral galaxy.}, which is determined in part by its overall mass, of which the bulge mass (which includes the central black hole)\endnote{To learn how the rotational velocity of a late-type galaxy correlates with the central black hole mass, see \citet{2019ApJ...877...64D}.} constitutes a proportional contribution according to the bulge/total light ratio of the galaxy.
    Verily, the lower surface brightness galaxies of our sample are likely to be less massive overall and less bulge-dominated.
    The difference in Hubble classifications between the two samples confirms this expectation.
    
Our sample is derived from the \citetalias{2011ApJS..197...21H} sample, which includes a total of 214 galaxies within $V_{\rm C} = 3.37$ $\times$ $10^4$\,$h_{67.77}^{-3}$\,Mpc$^3$.
Our study focuses on the dimmest 74/214 (34.6\%) galaxies among the sample.
Indeed, the dimmest 74 galaxies are morphologically distinct from its brighter counterpart of 140 galaxies.
The average galaxy from the dim sample is a Hubble subtype Sc (Hubble stage, $T=5.4\pm1.7$), whereas the average bright galaxy is Hubble subtype Sbc ($T=3.8\pm1.6$).
Which, if we consider the two distributions of Hubble stages as normal distributions, the agreement (overlapping area) between the two normal probability distributions is 63.4\%.
Thus, there is a slight agreement around the mean $T=4.4\pm1.8$ of the combined 214 galaxies, with the noticeable disagreement of 36.6\% stemming from the consistent excesses of $T\leq4$ stages for the bright sample and $T>4$ stages for the dim sample.


	While the galaxies of the faint sample are different in Hubble type, their distribution of arm number approximates that of \citet{2014ApJ...789..124D}.
	The dominant symmetry in both sets of galaxies is $m=2$ by far, with the faint sample showing an even higher percentage of two-armed spirals.
	There is also an excess of flocculent galaxies in the faint sample.
	These generally fainter galaxies have a tendency toward less organised structure, with symmetry either unclear, having a low arm-interarm contrast, or being more difficult to identify with more spurs than grand-design arms.
	Assigning a proper arm count to these messy galaxies is difficult.
	For fainter galaxies, it may simply be easier to resolve a two-armed spiral than to distinguish the number of arms in a less organized structure.
	This high number of flocculent galaxies explains the relative excess of $m=2$ and missing $m>2$ modes in the faint sample.
	As flocculent galaxies do not often have a dominant symmetry mode assigned to them, these missing assignments at least partially account for the dearth of galaxies with higher symmetry modes than $m=2$ and the corresponding excess of $m=2$ galaxies in the faint sample.


\section{Analysis}\label{sec:Analysis}

\subsection{Galactic Spiral-arm Pitch Angle Measurement}
	\subsubsection{\textsc{2dfft}}	
	
	Measurement of galactic logarithmic spiral-arm pitch angle is achieved through utilization of a modified two-dimensional fast Fourier transform code called \textsc{2dfft} \citep{1992A&AS...93..469P,1994A&AS..108...41S,2000A&A...359..932P,Seigar:2008,2012ApJS..199...33D,2016ascl.soft08015D}.
	This package decomposes digital images of spiral galaxies into superpositions of spirals of various pitch angles and number of arms, or harmonic modality ($m$).\endnote{
	There is not a way to classify a galaxy as, for example, ``three arms, plus or minus one arm.''
	However, other software such as \textsc{SpiralArmCount} \citep{Spirality,Shields:2015} is capable of measuring the relative contributions from each choice of the number of arms.
	Moreover, the \textsc{2dfft} software \citep{2012ApJS..199...33D,2016ascl.soft08015D} used in our current manuscript is capable of computing the pitch angle for the superposition of any number of arms.
	As such, it can report the pitch angle for the superposition of $m = 2$ \& 3, as well as $m = 2$, 3, \& 4, \textit{etc.}
	}
	Measurement using \textsc{2dfft} is most effective when applied to face-on galaxies; galaxies at higher inclination to line-of-sight are deprojected to a face-on orientation in pre-processing.
	Position and inclination angles are measured with the \textsc{iraf} \citep{Tody:1986,IRAF,1999ascl.soft11002N} routine \textsc{ellipse} \citep{1987MNRAS.226..747J}.
	We assume circular outer isophotes for our galaxy deprojections.
	Small errors in deprojection ($<$$10\degr$ of inclination) do not greatly alter the measurement of a mean pitch angle \citep{2012ApJS..199...33D}.
	However, departures from circularity of disks in these galaxies may increase the standard deviation for pitch angle measurements.
		
\begin{figure}[h]
	\begin{adjustwidth}{-4cm}{0cm}
	\centering
		\includegraphics[clip=true, trim= 1mm 1mm 1mm 3mm, width=18.46 cm]{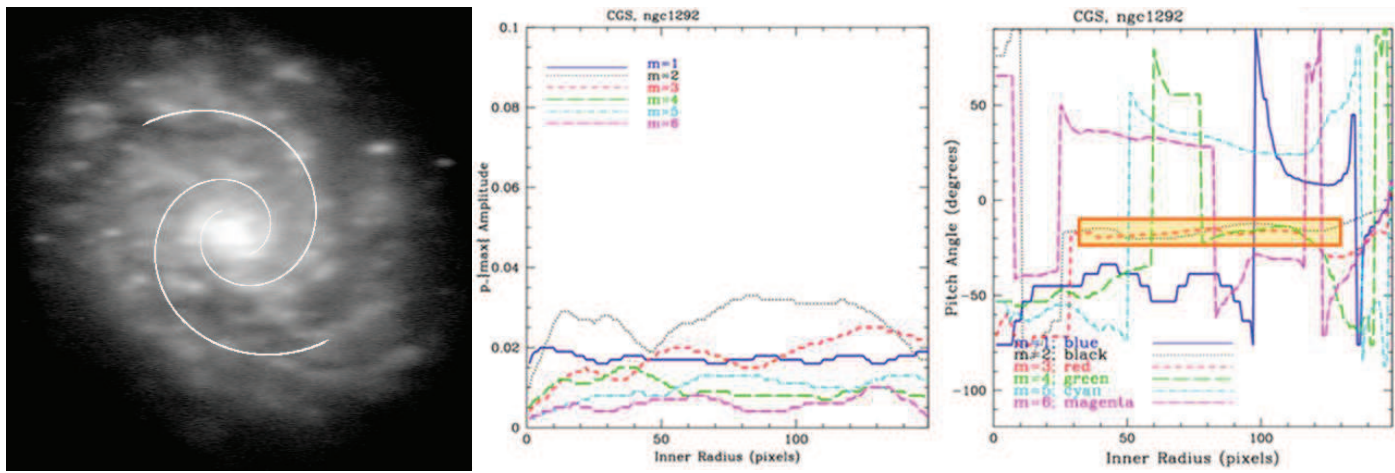}
		\end{adjustwidth}
		\caption{
		Example of spiral-arm pitch angle measurement for galaxy NGC~1292 using \textsc{2dfft}.
		\textbf{(Left)} Deprojected galaxy image in the $B$-band, with the best-fit pitch model ($m=2$, $\phi=-14\fdg87$) overlaid.
		\emph{N\!\!B}, we utilized the \textsc{SpiralArmCount} software \citep{Spirality,Shields:2015} to overlay the best-fit pitch model output from \textsc{2dfft}.
		Thus, the overlaid spiral is a representation of, and equivalent to, the inverse Fourier transform from \textsc{2dfft}.
		\textbf{(Middle)} Amplitude of each Fourier mode, with the $m=2$ symmetry (\textcolor{black}{{\hdashrule[0.35ex]{8mm}{0.4mm}{1pt}}}) being the most prominent for this galaxy.
		\textbf{(Right)} \textsc{2dfft} results, with the user-selected region of stable pitch angle measurement highlighted with an orange rectangle, yielding $\phi=-14\fdg87\pm3\fdg37$ for $m=2$ (\textcolor{black}{{\hdashrule[0.35ex]{8mm}{0.4mm}{1pt}}}).
		\emph{N\!\!B}, the stable region for the $m=2$ mode also shows stability for some less dominant, higher orders of symmetry, in this case $m=3$ and $m=4$.
		The two-arm symmetry is distinguishable in the image, though this flocculent galaxy blends the $m=2$ spiral with arm segments in different symmetry configurations.
		Particularly, images with higher signal-to-noise ratios will exhibit higher levels of agreement among multiple harmonic modes.
		Moreover, there will be less confusion over the chirality (i.e., clockwise or counterclockwise winding) of the galaxy.
		By visual inspection of the image in the left panel, we can safely ignore all of the positive pitch angle measurements displayed in the right panel. 
		}
		\label{fig:example}
	\end{figure}

	Within the \textsc{2dfft} routine, the center of the galaxy and an outer radius are specified by the user, and pitch angles are measured in annuli with an inner and outer radius.
	The inner radius is iterated from zero by integer pixel radii values to one pixel less than the user-selected outer radius of the galaxy.
	This produces a set of measurements all with the same outer radius but of varying inner radii.
	Often, the largest annuli will contain the core of the galaxy and any bar features present, results in a biased measurement for pitch angle.
	An inner radius of measurement, outside of which the pitch angle is relatively stable (ideally a large fraction of the total galactic radius) is selected by the user (see Figure~\ref{fig:example}).
	
	On the other end of the measurement, the smallest annuli are in the lowest surface brightness outer regions of the galaxy where signal-to-noise ratios are lower.
	These annuli also contain less of the rotation of the spiral as the annulus shrinks.
	These small outer annuli with unstable pitch angles are excluded.
	As a result, the measurements with small inner radii or large inner radii compared to the galaxy radius are excluded, while a region of medium annuli with stable pitch angle measurements is used to find the dominant pitch angle in the galaxy.
	
	The stable range of annuli should be characterized by a small variation in the measured value for pitch angle.
	We report the final pitch angle as the mean pitch angle within this stable region and the uncertainty in pitch angle is dependent on the standard deviation of the pitch angle within the stable region as well as the size of the stable region compared to the size of the galaxy.
	\textsc{2dfft} also returns the amplitude of each of the first six non-zero harmonic modes (\textit{i.e.}, $1\leq m \leq6$).\endnote{
	The software does compute $m=0$, however, this is the trivial solution because $\tan(\phi) = m/p_{\max}$, (where $p_{\max}$ is the Fourier frequency with the maximum amplitude) and thus $\phi\equiv0$ for $m=0$.
	The software is also capable of computing $m>6$, but these higher modes become redundant because they reflect the same trends seen in the modes $1\leq m\leq6$ when $m$ is divisible by 1, 2, 3, 4, 5, or 6. 
	}
	The dominant mode (number of arms) corresponds to the frequency with the maximum Fourier amplitude.
	Full details of our methodology for measuring pitch angles may be found in \citet{2012ApJS..199...33D}.
	
	\subsubsection{Use of \textsc{galfit} in Galaxies with Low Arm-interarm Contrast}\label{sec:galfit}
		
	\textsc{galfit} (version 3.0) is a two-dimensional fitting algorithm for galaxies, as described in \citet{2010AJ....139.2097P}.
	In addition to traditional bulge/disk fitting, the up-to-date version of \textsc{galfit} allows detailed fitting of complex and non-axisymmetric features in galaxies such as spiral arms, bars, rings, tidal tails, \textit{etc.}
	There are two advantageous approaches in utilizing \textsc{galfit} to assist in pitch angle measurement for a given galaxy.
\begin{enumerate}
	\item \textbf{Fully modeling the galaxy in question, including spiral structure, and extract the pitch angle of the spirals from the model.}
	The main advantage in this case is foreground stars and clumps in the galaxy structure go unmodeled and instead the focus is on reproducing radial light profiles and measurable spiral arms.
	A disadvantage to model fitting of spiral arms with \textsc{galfit} is the time cost per galaxy to fit a reasonable model.
	Spiral arms produced through \textsc{galfit} modeling behave a bit differently from real arms and may be difficult to match to real galaxy features.
	When working with large samples of galaxies, basic radial fitting without non-axisymmetric components is the preferred usage, and is the method employed for all galaxies in this sample.
	
	\item \textbf{Incorporating \textsc{galfit} models into pitch angle measurement to remove the radial light profile of the galaxy from the image before running \textsc{2dfft}.}
	This option offers a more practical route than full modeling.
	This technique is rapid to implement, and as a consequence is not prohibitively time intensive for measuring large samples of galaxies.
	Generally, a one- or two-component S{\'e}rsic profile \citep{Sersic:1963,Sersic:1968,Graham:2005} is fit to the target galaxy and removed, where the residual image primarily retains the spiral structure.
	
\end{enumerate}

\begin{figure}
		\begin{adjustwidth}{-4cm}{0cm}
		\centering
			\includegraphics[clip=true, trim= 0mm 0mm 0mm 0mm, width=18.46 cm]{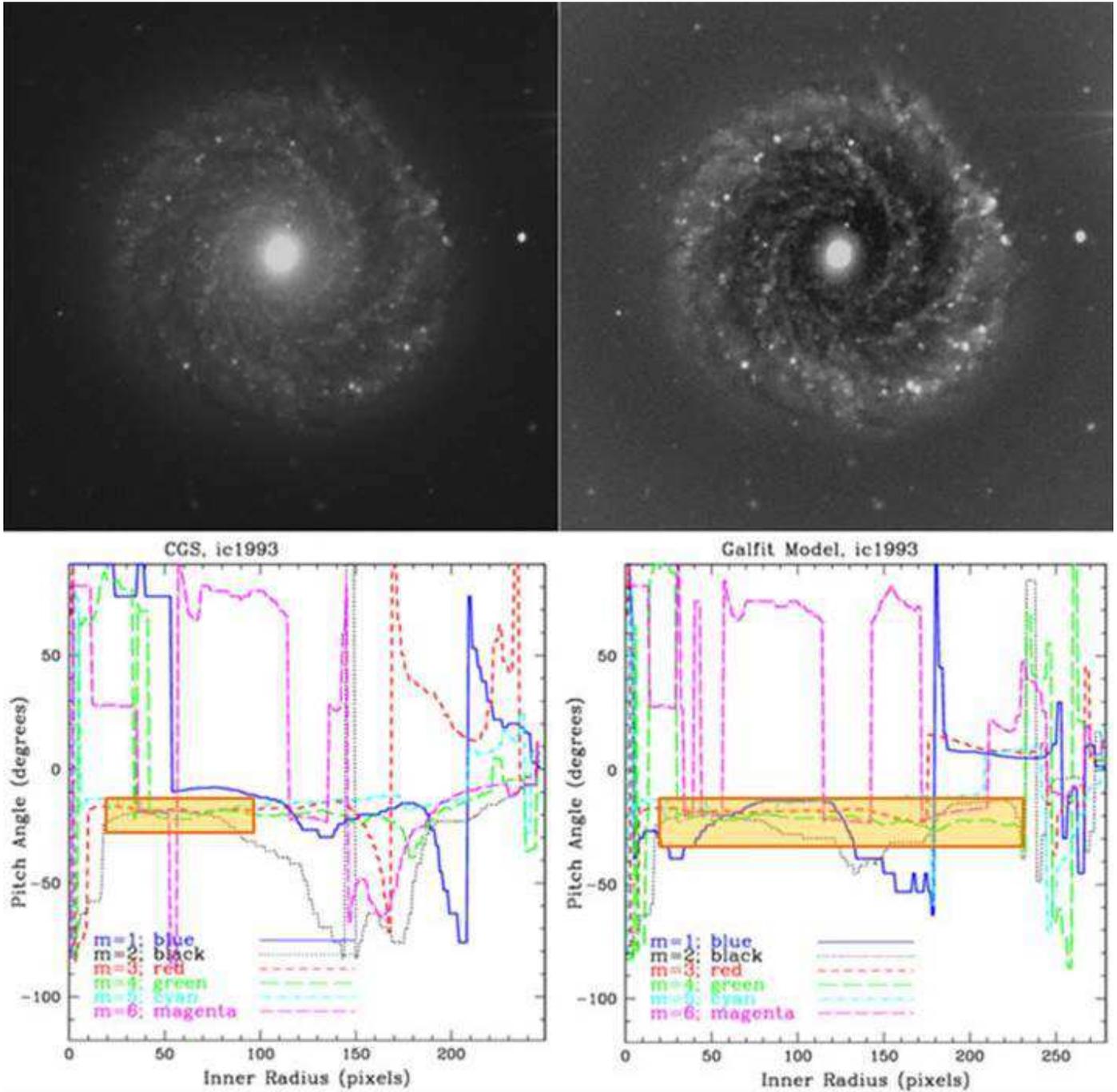}	
			\end{adjustwidth}
			\caption{
			Improved \textsc{2dfft} measurement of $m=2$ spiral-arm pitch angle (\textcolor{black}{{\hdashrule[0.35ex]{8mm}{0.4mm}{1pt}}}) for IC~1993 by means of \textsc{galfit} \citep{2010AJ....139.2097P}. 
			\textbf{(Left Column)} Original $B$-band image and corresponding pitch angle measurement.
			\textbf{(Right Column)} Modified image using \textsc{galfit} to remove a S{\'e}rsic profile from the galaxy, amplifying the visibility of the spiral arms.
			\textsc{2dfft} is capable of pitch angle measurement for both images; the user-selected stable region (highlighted orange rectangle) for the \textsc{galfit}-corrected image covers a larger range of galactic radii with a smaller standard deviation, thus producing a lower measurement uncertainty ($m=2$, $\phi=-20\fdg46\pm2\fdg48$).
			}
			\label{fig:galfitexample}
		\end{figure}

	The latter option of fitting a S{\'e}rsic profile to each galaxy is especially helpful in increasing the relative strength of spiral features at outer radii, as much of the surface brightness of the galaxy is removed by subtracting the model.
	We use this method of fitting and subtracting a S{\'e}rsic profile from each galaxy solely as a tool to enhance the spiral structure and benefit pitch angle measurements.
	Without careful multi-component decomposition \citep[\textit{e.g.},][]{Davis:2018,2019ApJ...873...85D,Sahu:2019a,Hon:2022}, simple S{\'e}rsic profiles fit to spiral galaxies generally overestimate the total flux by 15\% \citep{Sonnenfeld:2021}.
	Thus, the photometry of our S{\'e}rsic components are not pursued for this work.
	Galaxies before \textsc{galfit} light profile removal may be considered as ``center-weighted'' as most of the brightness of the galaxy is centrally located.
	By flattening the radial light profile of the galaxy with \textsc{galfit}, subsequent pitch angle measurements with \textsc{2dfft} tend to be more stable and reach higher galactic radii (see Figure~\ref{fig:galfitexample}) as opposed to favoring the inner radii of a given annulus.
	By employing S{\'e}rsic profile subtraction, we obtain more and better pitch angle measurements than would be possible otherwise.

		\subsubsection{Symmetrical Components}\label{sec:sym}
			
	Another technique for enhancing the accuracy of pitch angle measurements is to extract the symmetrical components of the target galaxy.
	Such a symmetrical-component image is created by dividing the original image into pixel pairs/groups, co-radial from the center of the galaxy and symmetrically distributed in rotation angle \citep{1992ApJS...79...37E,2012ApJS..199...33D}.
	Pairs/groups then adopt the lowest pixel value among them as the value for all of them.
	For example, for $m=2$ symmetry, all points radially opposed from one another at $180\degr$ adopt the lowest pixel value of each pair; for $m=4$, the lowest pixel value of the groups of four pixels $90\degr$ apart, and so on.
	The algorithm for performing the symmetric component isolation for any value of $m$ is given in \citet[][Equation~8]{2012ApJS..199...33D}, which is an extension of the original operation presented in \citet{Elmegreen:1992}.
    
\begin{figure}
\begin{adjustwidth}{-4cm}{0cm}
\centering
	\includegraphics[clip=true, trim= 0mm 0mm 0mm 0mm, width=18.46 cm]{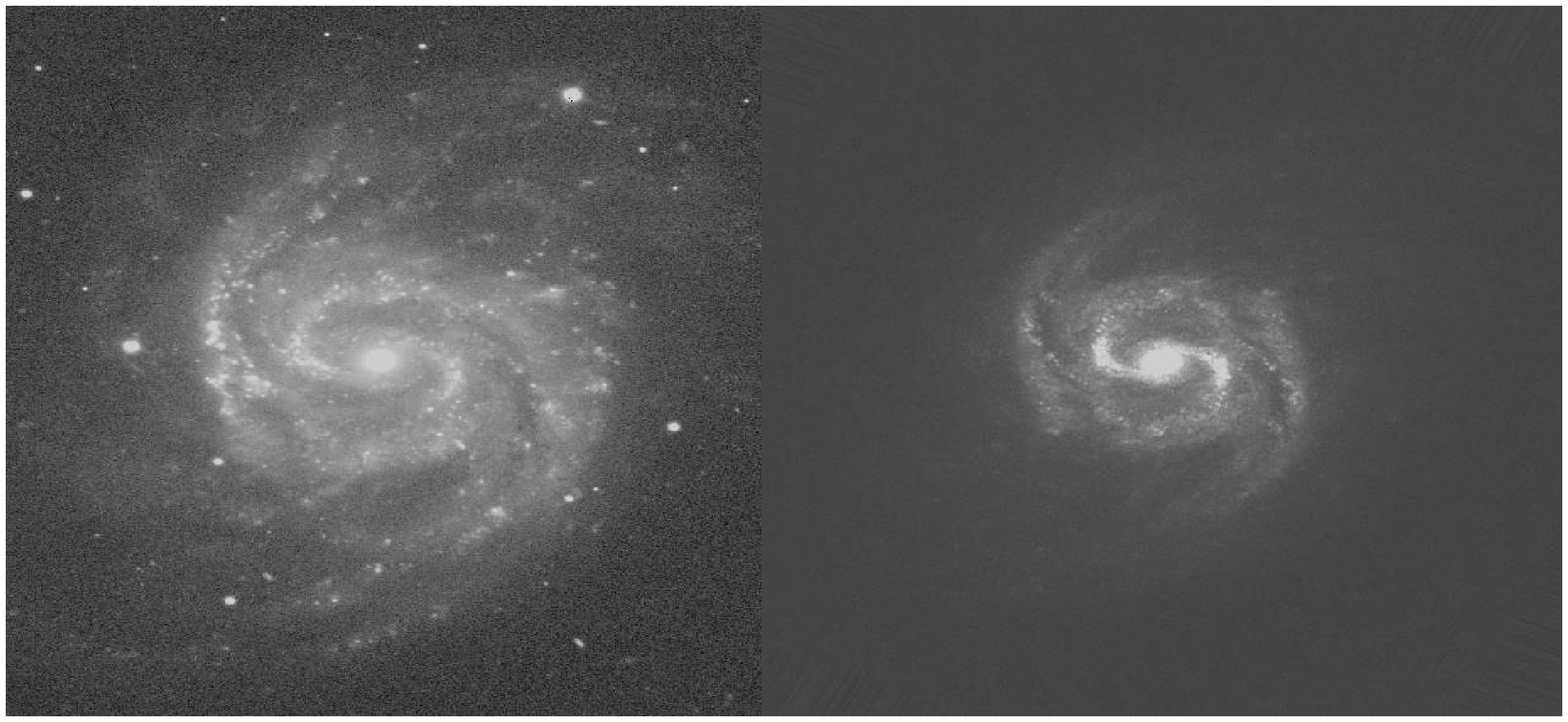}
	\end{adjustwidth}
	\caption{
	$m=2$ symmetrical component of galaxy NGC~1042.
	\textbf{(Left)} Original $B$-band image.
	\textbf{(Right)} Symmetrical-component image.
	Foreground stars and asymmetric bright features are removed and symmetrical features remain, allowing for more accurate measurement of spiral-arm pitch angles.
	Taking symmetrical components, where applicable, is a more rapid process for improving measurement than other methods, such as star subtraction.
	However, this comes at a cost of loss of information about the galaxy, as arm segments which do not fit the symmetry of the selected symmetrical component are removed.
	}
	\label{fig:symcompex}
\end{figure}

	Symmetric image preprocessing is applicable to galaxies where the arm symmetry is apparent, but where foreground stars or structural irregularities cause inaccurate measurements of the overall behavior.
	Although, taking symmetrical components is a double-edged sword, with tension between two characteristics: effective resolution and structural irregularities.
	Symmetric isolation tends to not only remove or reduce the effects of error-inducing features such as foreground stars, minor arm spurs, and flocculence from the image (see Figure~\ref{fig:symcompex}), but also removes detail from the galaxy.
	The effective pixel count of the galaxy is reduced by approximately a factor of $m$, as information is lost through symmetry folding.
	In most cases, this is a worthwhile trade-off of lost resolution for improved measurement of symmetrical features with lower associated error.
	
	For galaxies with many foreground stars, a full star subtraction using \textsc{daophot} \citep{1987PASP...99..191S} in \textsc{iraf} or similar programs may be advisable before producing symmetrical-component images for final measurement.
	Both \textsc{galfit} model subtractions and symmetrical-component images may be introduced individually or in sequence to galaxy images.
	Thus far, the most stable pitch angle measurements are obtained by first removing the S{\'e}rsic component and subsequently creating a symmetrical-component image of the residual from the S{\'e}rsic profile fit.
	Applying these techniques in tandem often increases the measurable radii range for the galaxy by 10\%--25\% of the galactic radius and reduces the uncertainty in pitch angle measurement for the galaxy by an average of $\approx$2$\degr$ in our error assignment.
	Indeed, all of the post-processing techniques mentioned in \S\ref{sec:galfit}--\S\ref{sec:sym} involve throwing away information from the image of a galaxy.
	Therefore, we do not apply these methods as a default; we always initially measure the unadulterated images and determine on a case-by-case basis whether one or more forms of post-processing is needed.

\subsection{Properties and Measurement of the Faint Sample}

Any comparison between two related populations of data requires an examination of which data, if any, is likely to be missing from each population.
When the \citet{2014ApJ...789..124D} sample was assembled, it was chosen for reasons of sample completeness, namely, galaxies within its distance and luminosity bounds were selected (bright enough to be observed and identified as spirals).
We relax brightness condition for the faint sample, but partially recover completeness by statistically accounting for missing galaxies.

There presumably exist many galaxies within the distance and luminosity limits of the faint sample which are not observed due to the brightness limits ($B_{\rm T}<12.9$\,mag) of the \citetalias{2011ApJS..197...21H}.
This subsample of galaxies which are too faint to be detected in the \citetalias{2011ApJS..197...21H} are missing from our population and need to be accounted for.
In addition, we fail to measure pitch angles for some fraction ($19/74\approx26\%$)\endnote{Since random inclination is the main limiting factor that prevents pitch angle measurement in spiral galaxies, this successful fraction of galaxies measured here is akin to measuring all galaxies with inclination angles less than $\lesssim$$75\degr$ (\textit{i.e.}, $55/74\approx74\%\approx1-\cos{75\degr}$), which is similar to the success rates in previous works \citep[\textit{e.g.},][]{2012ApJS..199...33D,2014ApJ...789..124D}.}\label{en:i} of the sample, with spirals classified as Magellanic spiral (Sm type) going completely unmeasured in the faint sample.\endnote{However, \citet{2014ApJ...789..124D} did manage to measure the pitch angles of three SBm type galaxies in their sample of 140 spiral galaxies.}
The faint sample is morphologically different from that of \citet{2014ApJ...789..124D}, containing an excess of Sc, Scd, Sd, and Sm types of galaxies and a dearth of Sa, Sab, Sb, and Sbc types.
Hubble types Sa and Sab are nearly absent from the faint sample (see Figure~\ref{fig:types}).

\begin{figure}
    \begin{adjustwidth}{-4cm}{0cm}
    \centering
	\includegraphics[clip=true, trim= 1mm 1mm 0mm 7mm, width=18.46 cm]{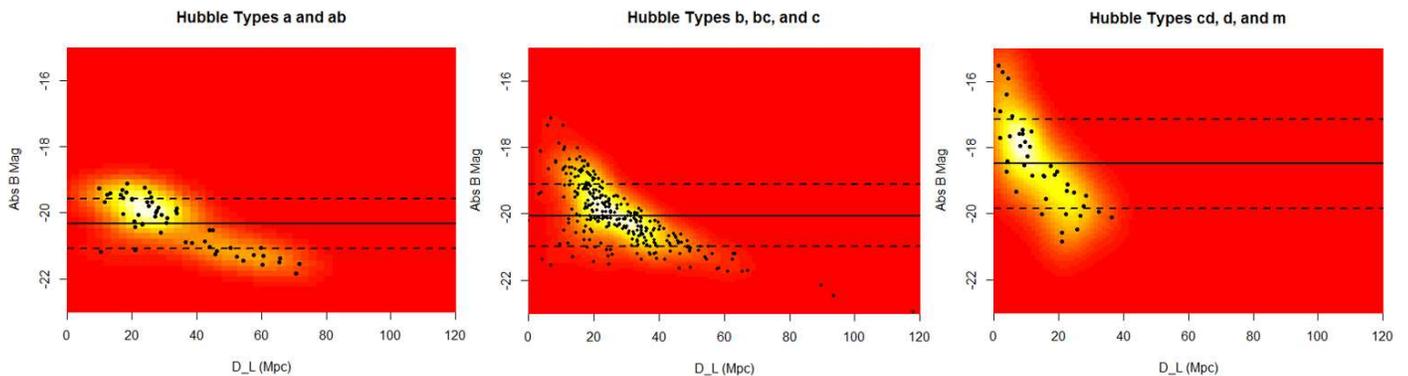}
	\end{adjustwidth}
	\caption{
	$\mathfrak{M}_{B}$ \textit{vs.} $D_{\mathrm{L}}$ diagram with mean absolute $\mathfrak{M}_{B}$ (solid horizontal lines) and standard deviation (dashed horizontal lines) of the full \citetalias{2011ApJS..197...21H} sample, sorted by Hubble type.
	Hubble types Sa and Sab spirals \textbf{(left panel)}, as reported in the \citetalias{2011ApJS..197...21H}, are confined to roughly $\mathfrak{M}_{B}=-19$\,mag and brighter, with a mean near $\mathfrak{M}_{B}=-20.5$\,mag.
	Galaxies identified as Sb, Sbc, and Sc \textbf{(middle panel)} span nearly the whole sample in magnitude.
	Galaxies identified as Sc, Scd, and Sm \textbf{(right panel)} have a mean near $\mathfrak{M}_{B}=-18.5$\,mag.
	It is expected that galaxies with later Hubble types have lower surface brightness, especially Sm dwarfs.
	Galaxy types through Sc, in the \citetalias{2011ApJS..197...21H}, all have similar mean $\mathfrak{M}_{B}$, while the latest types begin to deviate toward lower brightness.
	These heat maps illustrate increased clustering of data points with stronger hues of yellow.
	}
	\label{magbytype}
\end{figure}

\begin{figure}[h]
		\includegraphics[clip=true, trim= 0mm 0mm 1mm 0mm, width=1\columnwidth]{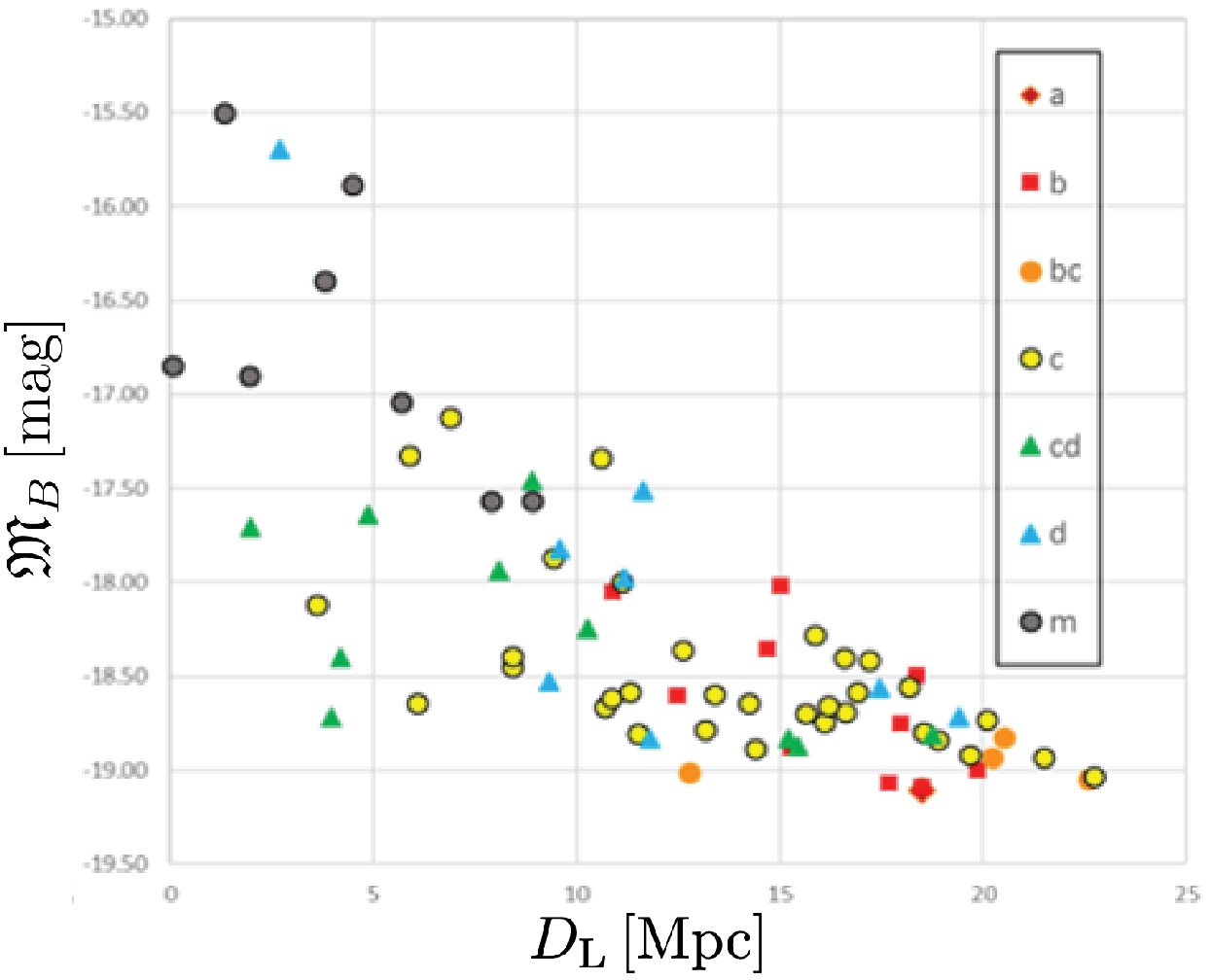}
		\caption{
		$\mathfrak{M}_{B}$ \textit{vs.} $D_{\mathrm{L}}$ diagram for just the faint sample, according to Hubble type.
		\emph{N\!\!B}, Sm dwarf Hubble types comprise nearly all the galaxies fainter than $\mathfrak{M}_{B}=-17$\,mag, and all of these galaxies went unmeasured in pitch angle (see also Note~\ref{en:i}).
		}
		\label{clusterandtypes}
	\end{figure}

We separate the \citetalias{2011ApJS..197...21H} by Hubble type into three groups to denote this populational difference in ($D_{\mathrm{L}}$, $\mathfrak{M}_{B}$)-space, see Figure~\ref{magbytype}, with Figure~\ref{clusterandtypes} showing the specific Hubble types for just the faint sample.
Figure~\ref{magbytype} illustrates the increased clustering (with stronger hues of yellow) of galaxies with broad grouping according to Hubble type in the $\mathfrak{M}_{B}$ \textit{vs.} $D_{\mathrm{L}}$ diagram for the entire \citetalias{2011ApJS..197...21H} sample.
Whereas, Figure~\ref{clusterandtypes} demarcates the different Hubble types for just our dim sample in the $\mathfrak{M}_{B}$ \textit{vs.} $D_{\mathrm{L}}$ diagram.
We find Sa and Sab galaxies appear at $\mathfrak{M}_{B}=-19$\,mag or brighter, with an average $\mathfrak{M}_{B}=-20.2$\,mag. 
Meanwhile, the Scd, Sd, and Sm subgroup populates a lower brightness regime, with an average $\mathfrak{M}_{B}=-18.5$\,mag.
Intermediate Hubble types span a large range of luminosities, but have an average $\mathfrak{M}_{B}$ consistent with the Sa and Sab group.
Looking at the pitch angle measurement success rates in the faint sample, we find that Sm types have a success rate of zero (all eight were not assigned a pitch angle), 67\% for Scd and Sd galaxies, and 85\% for Sa, Sab, Sb, Sbc, and Sc galaxies.
Even with Sm types removed, the Scd and Sd grouping shows a fainter average $\mathfrak{M}_{B}$.

	\subsection{Pearson Distribution Fitting of Pitch Angles}
	
\begin{figure}[h]
	\includegraphics[clip=true, trim= 1mm 4mm 6mm 0mm, width=\columnwidth]{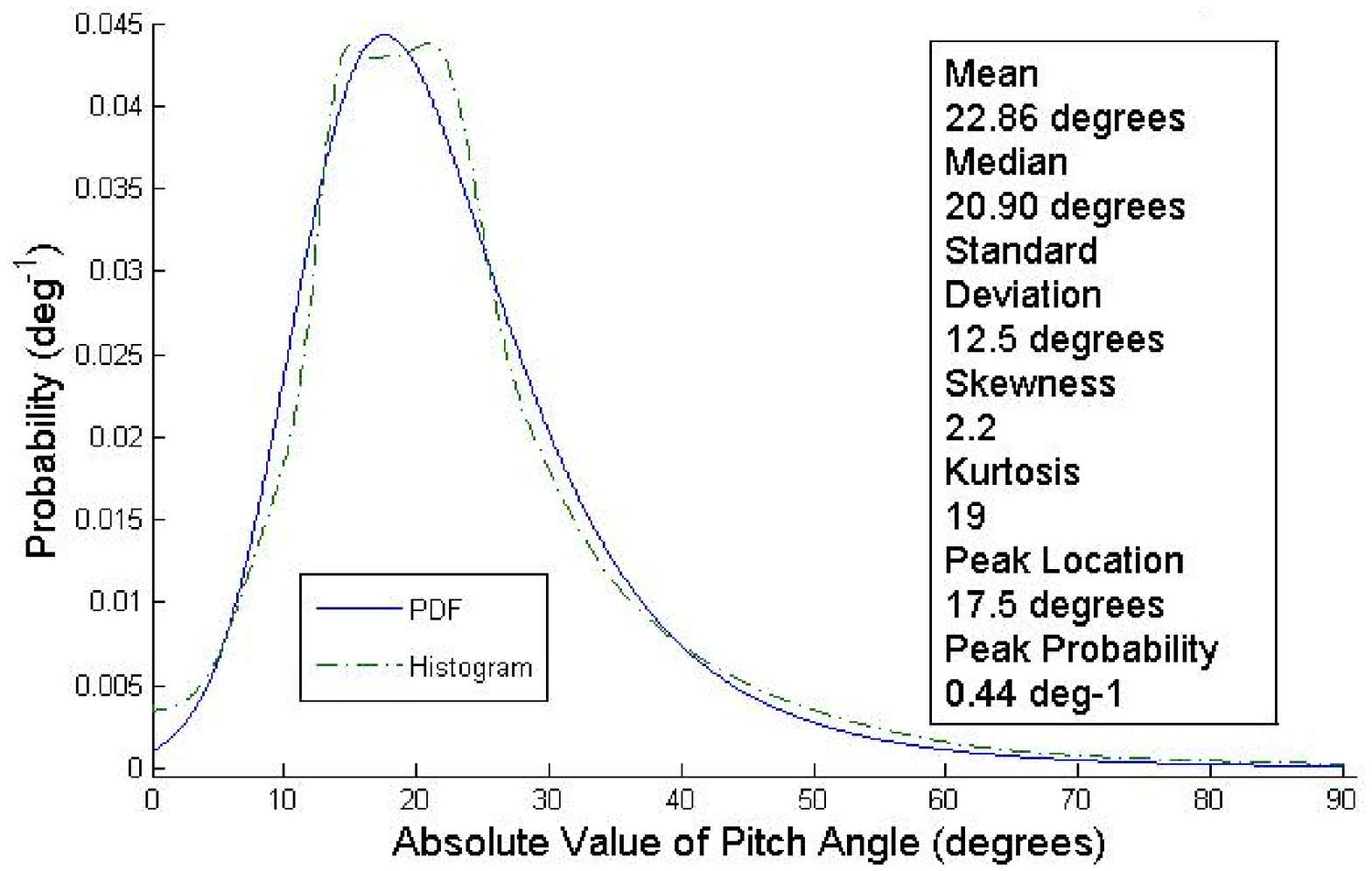}
	\centering
	\caption{
	Pitch angle distribution function (PADF) for the faint galaxies sample.
	The green, dashed-dotted-line distribution (\textcolor{MATLAB_green}{{\hdashrule[0.35ex]{1cm}{0.4mm}{1.5mm 3pt 0.5mm 2pt}}}) depicts the distribution of $|\phi|$, modeled as a ``binless'' histogram, where each data point is a Gaussian with mean equal to the measured pitch angle and standard deviation equal to its associated measurement uncertainty.
	The blue, solid-line distribution (\textcolor{MATLAB_blue}{{\hdashrule[0.35ex]{8mm}{0.4mm}{}}}) represents the probability density function (PDF) fit.
	Our best-fit skew-kurtotic-normal distribution to the data is shown.
	The faint galaxies sample shows higher skewness and kurtosis than its higher-luminosity counterpart \citep[\textit{cf.}][Figure~6]{2014ApJ...789..124D}, indicating data points with larger associated uncertainties and a heavier tail to the distribution.
	This graph was generated using the \textsc{r} package \textsc{pearson\_ds} \citep{R6}.
	}
	\label{pdist}
\end{figure}

	In \citet{2014ApJ...789..124D}, the probability density of pitch angles for their sample was calculated by creating a ``binless'' histogram which treats the dataset as a sum of Gaussians, with individual $|\phi|$ measurements and their error bars representing the means and standard deviations, respectively.
	This histogram was then statistically fit to a probability density function (PDF) based on the statistical properties of the dataset.
	This skew-kurtotic-normal fit of the PADF could then be transformed into a BHMF through the $M_{\bullet}$--$\phi$ relation \citep{2013ApJ...769..132B}.
	We reproduce these calculations for the current dataset (Figure~\ref{pdist}).
	
	Given the peak locations of the bright sample and faint galaxies sample match up reasonably well, the increased skew in the faint galaxies sample is interpreted mainly as a stronger tail of probability density at high (past-peak) $|\phi|$.
	Moreover, excess kurtosis in the faint galaxies sample is indicative of a heavy tail to the distribution.
	The excess density at the peak may be related to the excess of $m=2$ spirals in the faint sample.
	Galaxies with two-arm symmetry tend toward successful pitch angle measurement as compared to flocculent galaxies or galaxies with higher-order arm symmetries
	The two-fold symmetry galaxies are often more grand design-like, with higher arm-interarm contrast, whereas galaxies assigned $m=4$ or $m=6$ symmetries have a tendency toward clumpiness, with an agglomeration of many arm segments and spurs rather than sharply defined, contiguous arms.
	
	Difficulties associated with the measurement of (faint) flocculent galaxies in the sample may slightly bias the measurement toward the peak of the distribution.
	\citet{Elmegreen:1987} devised an arm classification system that aims to quantify the level of flocculence in spiral galaxies.
	Their system groups spiral galaxies into a dozen arm classes (AC), with AC 1--4 considered flocculent and AC 5--12 are grand design.
	Subsequent works \citep{2012ApJS..199...33D,Yu:2018} demonstrate that (although there is significant scatter) flocculent galaxies tend to have higher $|\phi|$ than grand design galaxies.
	Therefore, if we fail to measure flocculent galaxies at a higher rate than grand designs, then that would likely tend our overall PADF toward lower $|\phi|$ (and higher $M_{\bullet}$).

	Another possible cause of the peak excess for the faint sample is the measurement success rate by Hubble type, paired with the Hubble type abundance in the sample.
	As compared to the \citet{2014ApJ...789..124D} sample, ours has fewer Sa--Sbc Hubble types and an excess of Sc--Sm types.
	The Sc--Sm types more commonly evade pitch angle measurement.
	Also, the peak excess might be the result of the relative lack of Sa and Sab Hubble types in the faint sample.
	The absence of these higher-mass spiral galaxies from the distribution would naturally raise the probability densities estimated for all other masses.
	
	The combination of Sa and Sab types largely missing from the sample and Scd and Sd types more commonly evading pitch angle measurement might account for the lower probability densities on either side of the peak of the PADF, and a corresponding relative amplification of the peak.
	When combined with the other previously mentioned causes, it is not all that surprising that the fainter galaxy sample would show a higher peak probability than its brighter counterpart.
	Furthermore, the fundamental plane \citep{Davis:2015} would support the idea of smaller bulges of these later-type galaxies being compatible with tighter pitch angles (\textit{i.e.}, lower $|\phi|$) if they had lower gas densities in their disks.
	Additionally, a further complication to consider is the environment of a galaxy.
	For example, \citet{Bellhouse:2021} demonstrate that cluster galaxies exhibit ``unwinding'' of spiral arms (\textit{i.e.}, higher $|\phi|$) in their outer edges via ram-pressure stripping.
	Thus, this phenomenon works in opposition to the effect of decreased gas densities in disks.
	These effects are prospective subjects for further study.
	
\begin{figure}
		\begin{adjustwidth}{-4cm}{0cm}
		\centering
		\includegraphics[clip=true, trim= 0mm 0mm 2mm 4mm, width=18.46 cm]{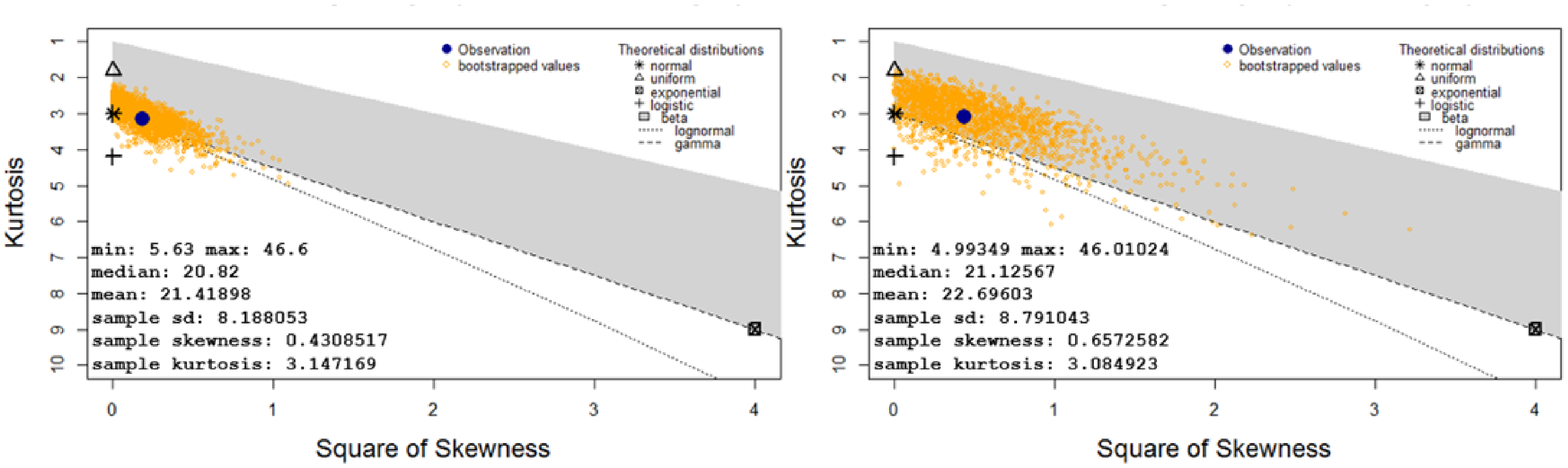}
		\end{adjustwidth}
		\caption{
		Comparison of Cullen and Frey graphs \citep{cullen1999probabilistic} with a 2000-sample bootstrap for the Pearson distributions of the \citet{2014ApJ...789..124D} sample \textbf{(left)} and faint galaxies sample \textbf{(right)}.
		Despite the Pearson distribution for the faint sample not fitting well to its data (\textit{i.e.}, not enough skewness or kurtosis), it is centered farther from the normal distribution on the Cullen and Frey graphs, illustrating this difference between the two samples.
		These plots were generated using the \textsc{r} package \textsc{fitdistrplus} \citep{R2}.
		}
		\label{cfcomb}
	\end{figure}

	Cullen and Frey graphs \citep{cullen1999probabilistic} are used to compare functions in skewness-kurtosis space.
	They are produced for each sample, using the \textsc{r} package \textsc{fitdistrplus} \citep{R2}, to further scrutinize the distribution shape in the context of these parametric solutions (see Figure~\ref{cfcomb}).
	Possible values of the skewness and kurtosis are calculated by bootstrapping (with a 2000-sample bootstrap) to illustrate the likely functional form of the parent distribution.
	The outcome is the bright sample more closely resembling a normal distribution than the faint galaxies sample.
	
	The excess skewness and kurtosis of the faint sample, as interpreted in this way, are somewhat reduced.
	The statistical fit of these values to the data did not match well to the binless histogram, and the scatter in bootstrapped values for the faint sample is much larger.
	Accordingly, differences retained between the samples at the tail of the distribution are not ruled out.
	Consequently, we perform non-parametric fitting of the samples to further compare their tail behavior.


\section{Pitch Angle Distribution Function}\label{sec:PADF}
	
	In order to assess whether the BHMF for the faint sample's lower luminosity regime of galaxies resembles those in preceding studies of brighter galaxies, an effective tool is to examine the PADF of these galaxies in comparison to prior work.
	If the current set of galaxies comprise the same population as their brighter counterparts, their PADFs will mirror one another, whereas if the current sample encompasses a second population of galaxies (presumably at the low-mass end of the distribution) this would alter the shape of the BHMF in this regime.
	
	\subsection{Non-parametric Fitting}
		
\begin{figure}
	\includegraphics[clip=true, trim= 0mm 5mm 8mm 18mm, width=\columnwidth]{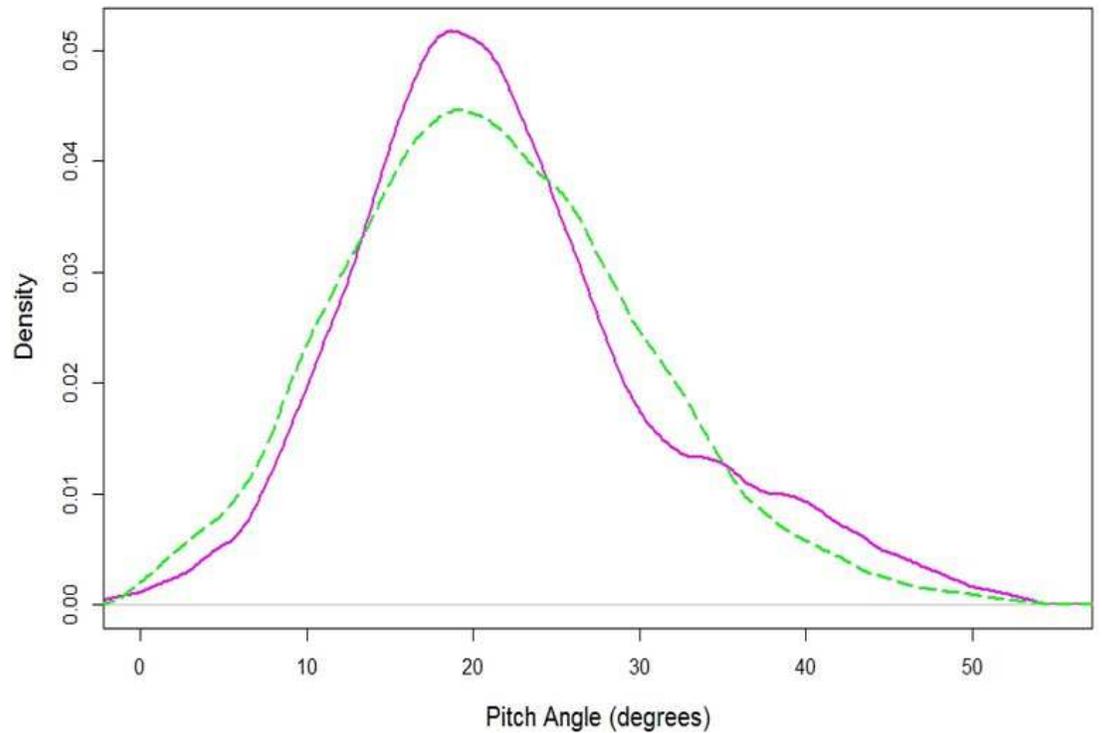}
	\caption{
	Comparison of initial kernel density estimator (KDE) fits of the probability densities for the \citet{2014ApJ...789..124D} bright sample (\textcolor{LimeGreen}{{\hdashrule[0.35ex]{9mm}{1pt}{1mm}}}) and our faint galaxies sample (\textcolor{magenta}{{\hdashrule[0.35ex]{9mm}{0.4mm}{}}}).
	The faint galaxies sample has a slightly higher peak and heavier tail than its counterpart.
	Both distributions are strikingly similar at the lower pitch angle (higher mass) end of the function.
	The two functions reach peak probability density at nearly the same pitch angle. 
	This graph was generated using the \textsc{r} package \textsc{kedd} \citep{R7}.
	}
	\label{KDEminevsben}
\end{figure}

	While parametric solutions to distribution fitting are simpler to express in functional forms, with insufficient data they are sometimes overly constrained.
	It is clear the initial Pearson fit to the faint sample was inadequate, likely due to the low number (55) of galaxies in the sample.
	In such a low-number case, a non-parametric kernel density estimator (KDE) approach is more appropriate for PDF fitting and dataset comparison (see Figure~\ref{KDEminevsben}).
	The distribution of the \citet{2014ApJ...789..124D} sample is closer to Gaussian, due to both morphology trends and to the larger sample size of their distribution.
	For a fair assessment, we utilized a KDE to fit both datasets.
	
	Even though the two samples are morphologically distinct, their pitch angle distributions are appreciably similar.
	A possible tendency towards very loose spirals at our distribution's tail betrays a sign of the shift to higher $|\phi|$ one may presume in a sample dominated by Sc and Sd Hubble types.
	This bump in the tail of the distribution is small, and measurement errors for \textsc{2dfft} tend to be larger in higher $|\phi|$ regimes.
	As such, further analysis (as follows) is required to ascertain the strength of the tail.

	KDE fits of the two distributions are performed as a method of comparing the samples.
	As with the parametric (Pearson distribution) fitting solutions, both samples have very similar peak pitch angles. 
	The bright sample shows a broader and lower peak with excess probability slightly past the peak, but a deficit probability far past the peak in the high pitch angle regime.
	This naïve Gaussian KDE, however, does not fully take into account some of the behaviors of the data which might affect the probability distribution.
	Among these factors to correct for are bandwidth selection in performing the KDE and weighting for heteroscedasticity in the data.
	Confidence intervals around the best fit KDE are also requisite to attain a comparison of samples that best shows their compatibility.

\subsection{$\phi$-dependent Errors and Bandwidth Optimization}

\subsubsection{Heteroscedasticity}

	Heteroscedasticity is an occurrence where the variance of a dataset changes for various subpopulations of the dataset.
	A common case of this is where the magnitude of the errors in one variable are covariant with the magnitude of another of the variables being examined.
	In order to properly the PDFs of the two samples using KDE, heteroscedasticity of the datasets needs to be assessed.
	The heteroscedasticity of pitch angle measurements was not analyzed in the methods paper of \citet{2012ApJS..199...33D}.
	We make an attempt here to address the issue, but leave a full accounting of the problem to subsequent work.

    In the case of galactic pitch angle measurement, recorded errors are expected to grow with pitch angle.
    Spiral galaxies with small pitch angles (tightly wound) tend to be measured with more accuracy than those with large pitch angles (loosely wound).
    One reason for this error growth with pitch angle is largely the result of the measurement technique employed.
    Galaxies with tight spirals have a broader swath of phase angles (more revolutions) of spiral arm to measure, while it is unlikely that loose spiral patterns will remain intact for even one measurable revolution of its galaxy.\endnote{
    Combining Equation~3 from \citet{Shields:2015} and Equation~3 from \citet{Davis:2017}, even a modest $\phi=20\degr$ spiral that remains intact for $2\pi$ radians and begins at an inner galactic radius of 1\,kpc would extend out to an outer radius of $\approx$10\,kpc and cover a total arc length of $\approx$26\,kpc as it wraps around the galaxy.}
    
\begin{figure}
    \begin{adjustwidth}{-4cm}{0cm}
		\centering
		\includegraphics[clip=true, trim= 0mm 0mm 0mm 0mm, width=9.18cm]{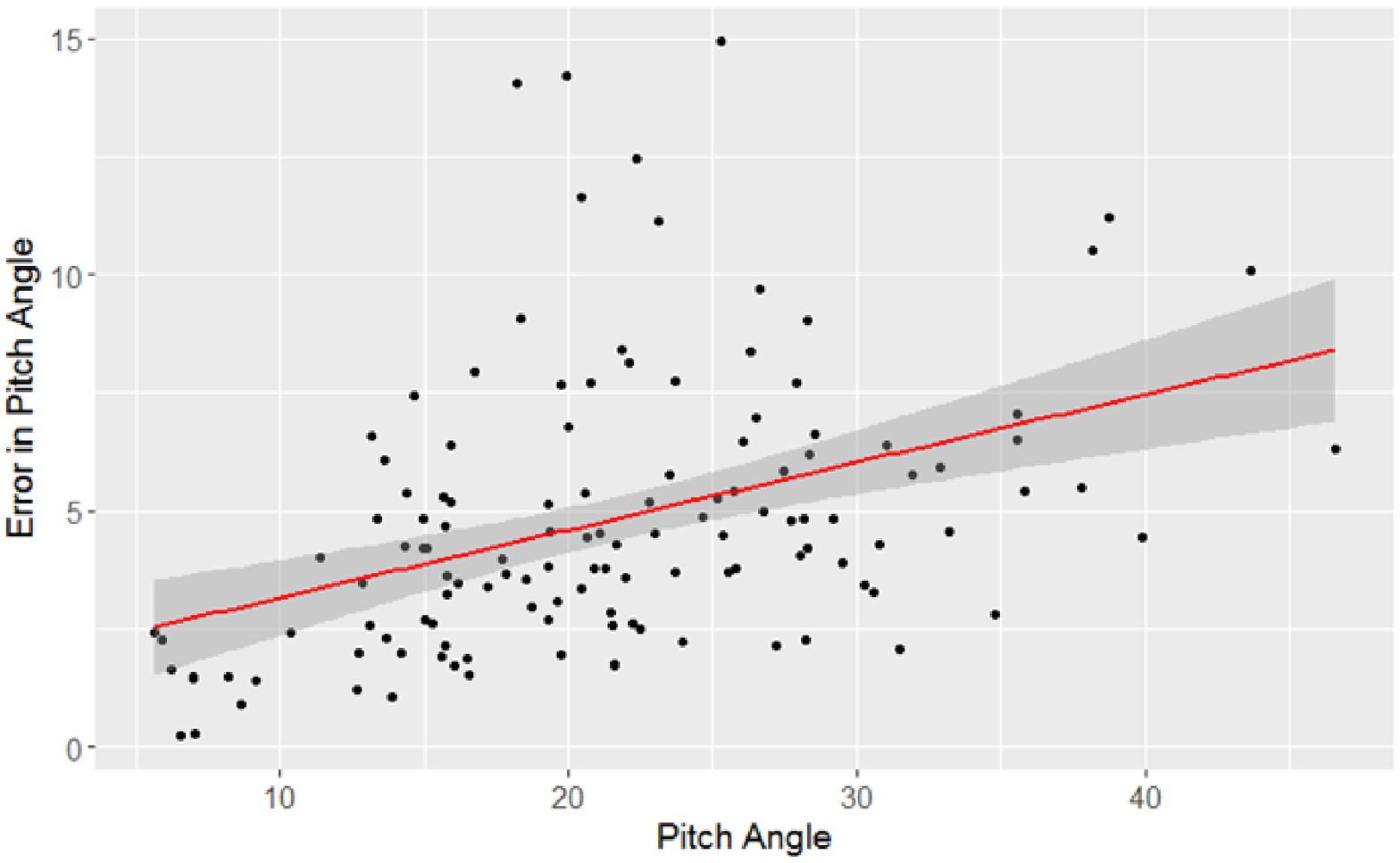}
		\includegraphics[clip=true, trim= 0mm 0mm 0mm 0mm, width=9.18cm]{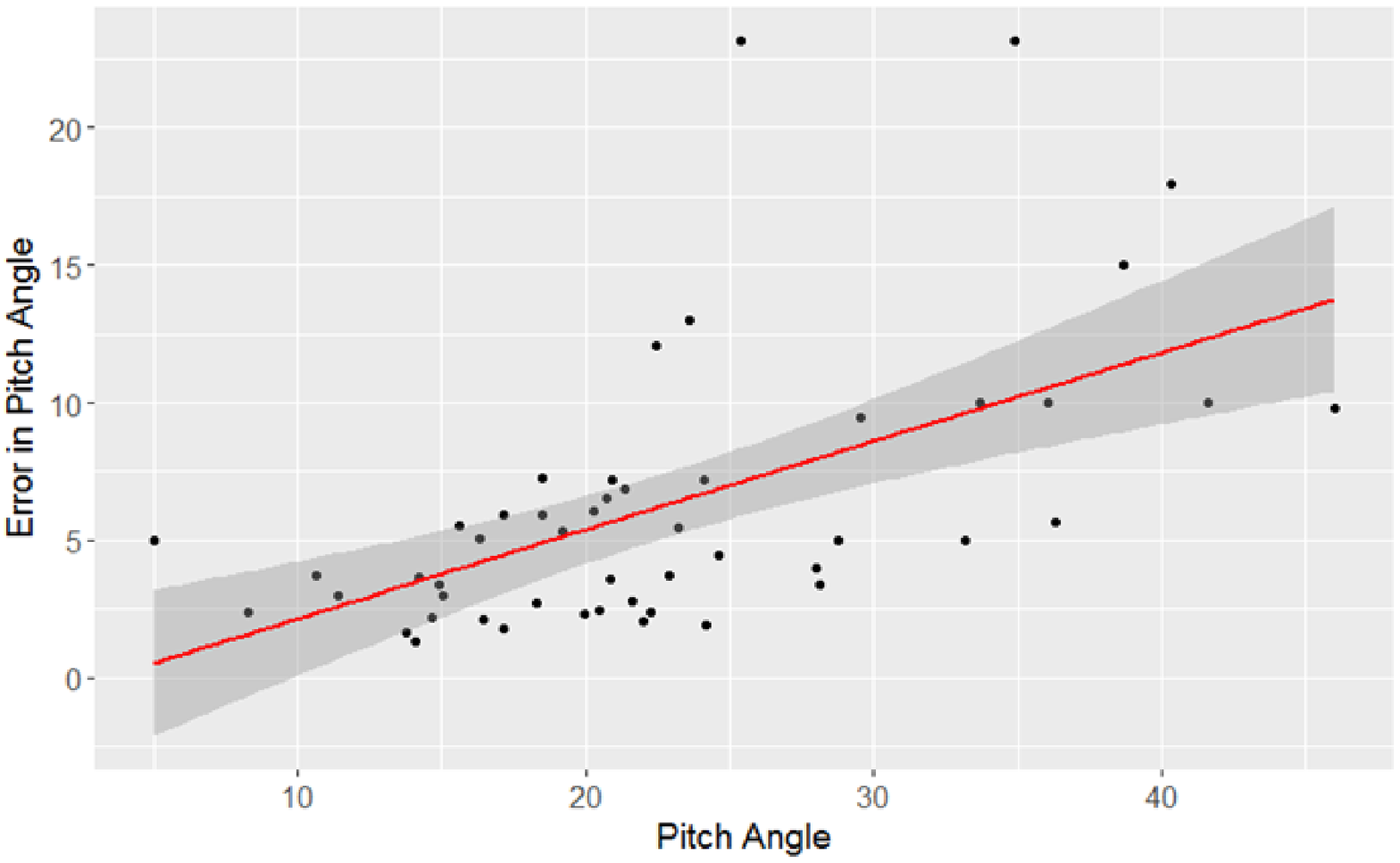}
		\end{adjustwidth}
		\caption{
		Measurements of pitch angle are heteroscedastic, \textit{i.e.}, $\delta\phi \propto|\phi|$.
		\textbf{(Left)} $\delta\phi$ (uncertainty in pitch angle measurement) as a function of $|\phi|$ for the \citet{2014ApJ...789..124D} sample.
		\textbf{(Right)} $\delta\phi$ as a function of $|\phi|$ for the faint galaxies sample.
		Linear fits of heteroscedasticity (\textcolor{red}{{\hdashrule[0.35ex]{8mm}{0.4mm}{}}}) include shaded 95\% confidence intervals.
		Both data sets show increasing measurement errors as a function of pitch angle absolute value.
		These graphs were generated via \textsc{r} scripts from \citet{R4}.
		}
		\label{heterosca}
	\end{figure}

    Figure~\ref{heterosca} shows that both samples do have uncertainties in pitch angle measurement ($\delta\phi$) which grow as a function of $|\phi|$, in accordance with expectations (\textit{i.e.}, $\delta\phi \propto|\phi|$).
    We fit a linear model fit to their relations.
    Robust regressions are applied in order to assign weights which minimize the impact of outliers to the fit.
    From the linear fits to the heteroscedasticity of the data, Huber weights \citep{Huber} are calculated, which are then applied to each galaxy in distribution fitting.
    In this way, the excess error on the tail of the distribution is subsequently accounted for in the determination of intrinsic probability density of the BHMF (see \S\ref{sec:AfH}).
    
	\section{Black Hole Mass Function}\label{sec:BHMF}
	
	The PADF is transformed into a BHMF using the $M_{\bullet}$--$\phi$ relation as described in \citet[][Equation~4]{2013ApJ...769..132B}:
	\begin{equation}
	\log(M_{\bullet}/\mathrm{M}_{{\sun}})=(8.21\pm0.16)-(0.062\pm0.009)|\phi|,
	\label{eqn:B+13}
	\end{equation}
	with a scatter of 0.38\,dex in black hole mass.
	Error propagation is performed before distribution fitting for the BHMF of each sample.
	Moreover, careful treatment of KDE produces updated estimates of the BHMF for both sets of spiral galaxies in the local Universe.
	The distribution of masses for the faint galaxies sample largely mirrors that of \citet{2014ApJ...789..124D}, with the anomalies mentioned earlier in the description of the PADF now translated to mass space, namely, the excess density at the peak (and somewhat for the tail for the faint sample) and under density of the faint sample at black hole masses higher than the mass of peak probability.
	
	\subsubsection{Bandwidth Selection}

\begin{figure}
	\includegraphics[clip=true, trim= 3mm 2mm 3mm 12mm, width=\columnwidth]{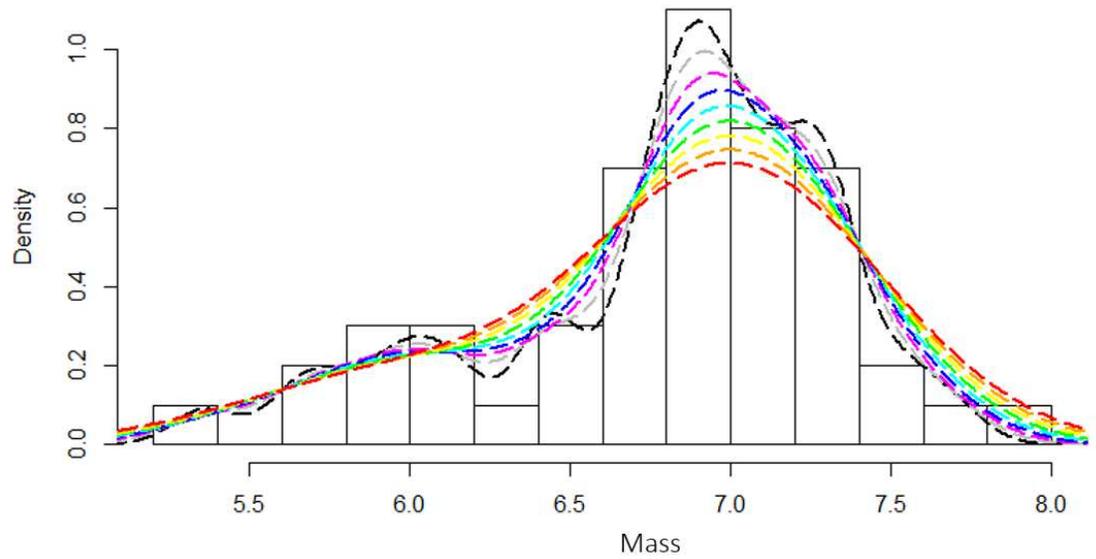}
	\caption{
	Bandwidth comparison for the faint sample's BHMF.
	At higher bandwidths, the faint sample more closely resembles the \citet{2014ApJ...789..124D} sample, while at lower bandwidths, some differences emerge.
	Bandwidth optimization techniques yield a bandwidth of 0.22\,dex (\textcolor{cyan}{{\hdashrule[0.35ex]{8mm}{1pt}{1mm}}}), in $\log(M_{\bullet}/\mathrm{M_{\sun}})$, as optimal, i.e., the smaller bandwidths are \emph{undersmoothed} and the higher bandwidths are \emph{oversmoothed}.
	Bandwidth selection does not strongly impact the height of the low-mass tail of the distribution, but more noticeably affects the peak density.
	Bandwidths of (0.10, 0.13, 0.16, 0.19, 0.22, 0.25, 0.28, 0.31, 0.34)\,dex are depicted as {\hdashrule[0.35ex]{8mm}{1pt}{1mm}}, \textcolor{gray}{{\hdashrule[0.35ex]{8mm}{1pt}{1mm}}}, \textcolor{magenta}{{\hdashrule[0.35ex]{8mm}{1pt}{1mm}}}, \textcolor{blue}{{\hdashrule[0.35ex]{8mm}{1pt}{1mm}}}, \textcolor{cyan}{{\hdashrule[0.35ex]{8mm}{1pt}{1mm}}}, \textcolor{green}{{\hdashrule[0.35ex]{8mm}{1pt}{1mm}}}, \textcolor{yellow}{{\hdashrule[0.35ex]{8mm}{1pt}{1mm}}}, \textcolor{orange}{{\hdashrule[0.35ex]{8mm}{1pt}{1mm}}}, and \textcolor{red}{{\hdashrule[0.35ex]{8mm}{1pt}{1mm}}}, respectively.
	A histogram with bin widths of 0.20\,dex (similar in width to the optimal bandwidth) is additionally shown in the background.
	}
	\label{m1_gauss}
\end{figure}

	Bandwidth selection is required to produce a representative KDE.
	We employ maximum likelihood cross validation (MLCV) to assess the optimal bandwidth for the data.
	The bandwidth in KDE may be thought of as a smoothing parameter of the fit when the number of observations is low.
	Higher bandwidths create functions with more smoothing while smaller bandwidths correspond to fits which are allowed more local variance.
	To avoid over or under smoothing the fit, we explored the effects of bandwidth selection on the behavior of the distribution.
	While the peak height of the faint sample varies with bandwidth, the tail behavior of the distribution is largely independent of bandwidth (see Figure~\ref{m1_gauss}).
	This means we can trust the tail behavior of interest to be similar regardless of the chosen bandwidth, though the MLCV bandwidth is likely to be the most representative of the sample.
	
\subsubsection{Accounting for Heteroscedasticity}\label{sec:AfH}

\begin{figure}
    \begin{adjustwidth}{-4cm}{0cm}
		\centering
	\includegraphics[clip=true, trim= 0mm 0mm 5mm 5mm, width=18.46 cm]{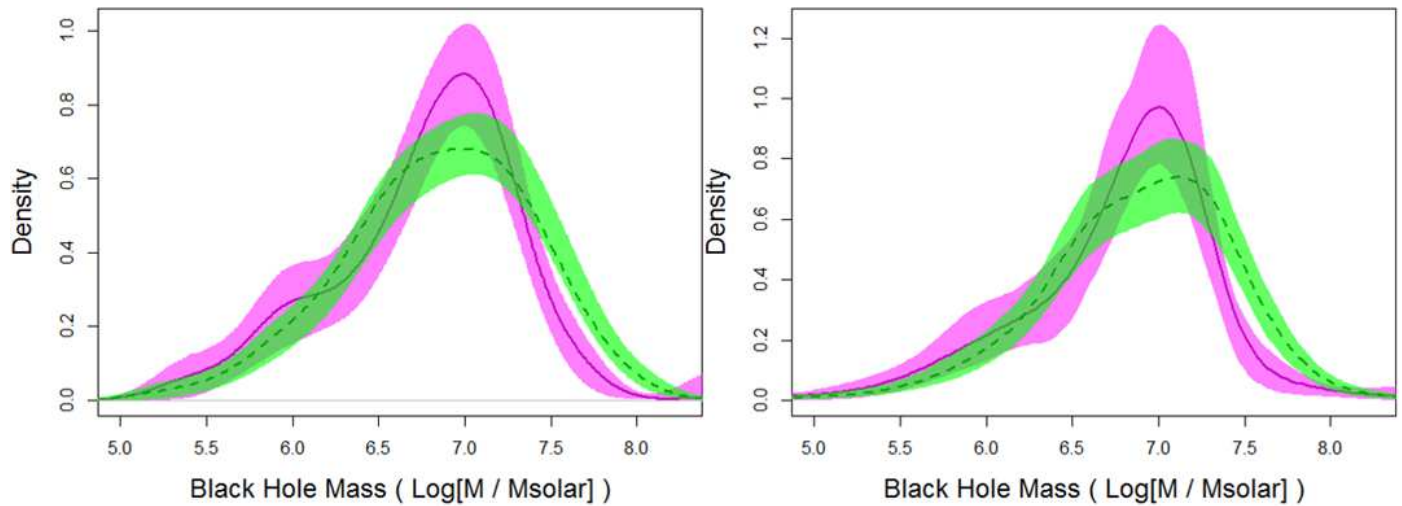}
	\end{adjustwidth}
	\caption{
	\textbf{(Left)} Faint sample (\textcolor{magenta}{{\hdashrule[0.35ex]{8mm}{0.4mm}{}}}) and \citet{2014ApJ...789..124D} sample (\textcolor{LimeGreen}{{\hdashrule[0.35ex]{8mm}{1pt}{1mm}}}) BHMFs as fit with bandwidth-selected KDE ($\text{bandwidth}=0.22$\,dex) and robust weighting.
	\textbf{(Right)} Adaptive bandwidth KDE BHMFs.
	90\% confidence intervals, found using 10,000-sample bootstrapping, are shown for each estimate.
	The faint sample shows a slight excess at the low-mass tail of the distribution, while the \citet{2014ApJ...789..124D} sample alternatively shows an excess at the high-mass end, which is likely due to more Sa and Sab type galaxies in their sample.
	}
	\label{static_adaptive}
\end{figure}

    Using our MLCV-generated bandwidth with our robust-weighted, heteroscedasticity-corrected samples, we convert from $|\phi|$ to $M_{\bullet}$ using the $M_{\bullet}$--$\phi$ relation and produce a KDE of the BHMF for the two samples (see Figure~\ref{static_adaptive}).
    Errors are assessed using bootstrapping to find the 90\% confidence interval around each density estimate.
    We also tested an alternate technique, adaptive bandwidth KDE, wherein the shape of the kernel used to produce the density is allowed to change shape (broaden or tighten) based on local properties of the data.
    Adaptive bandwidth techniques are advantageous when estimating density for heavy-tailed or multimodal distributions, such as our faint galaxies sample.
    Confidence intervals are assessed around each sample using a 10,000-sample bootstrap of the density estimation.
    
	\subsection{Expectation-maximisation Clustering Analysis}

	In order to further understand the behavior of the faint sample, we employed various clustering techniques.
	As the low-mass tail of the distribution of faint sample galaxies is heavy, and there is some indication of population differences between the faint sample and its brighter counterpart despite similar peak behavior, it is useful to explore whether the tail of the faint sample represents a distinct second population.
	Simple normality tests are not conclusive as to whether there is a true second population within the sample.
	Therefore, we perform expectation-maximization (EM) clustering analysis \citep{10.2307/2984875,10.1214/aos/1176346060} of the PADF to divide the faint sample into high-$|\phi|$ and low-$|\phi|$ subsamples.
	
	However, examining the coincidence of population sorting by Hubble types reveals the high-$|\phi|$ galaxies consist primarily of Sc, Scd, and Sd Hubble types.
	Not all of the Sc, Scd, and Sd types were sorted into the high-$|\phi|$ cluster, but none of the Sa--Sbc types were assigned to the high-$|\phi|$ cluster.
	An examination shows that some significant fraction of the very late Hubble types compose the second population.
	If the distribution is to be treated as one function rather than two, the single population carries a heavy tail made up of these latest Hubble types.

\begin{figure}
	\begin{adjustwidth}{-4cm}{0cm}
		\centering
	\includegraphics[width=9.18 cm]{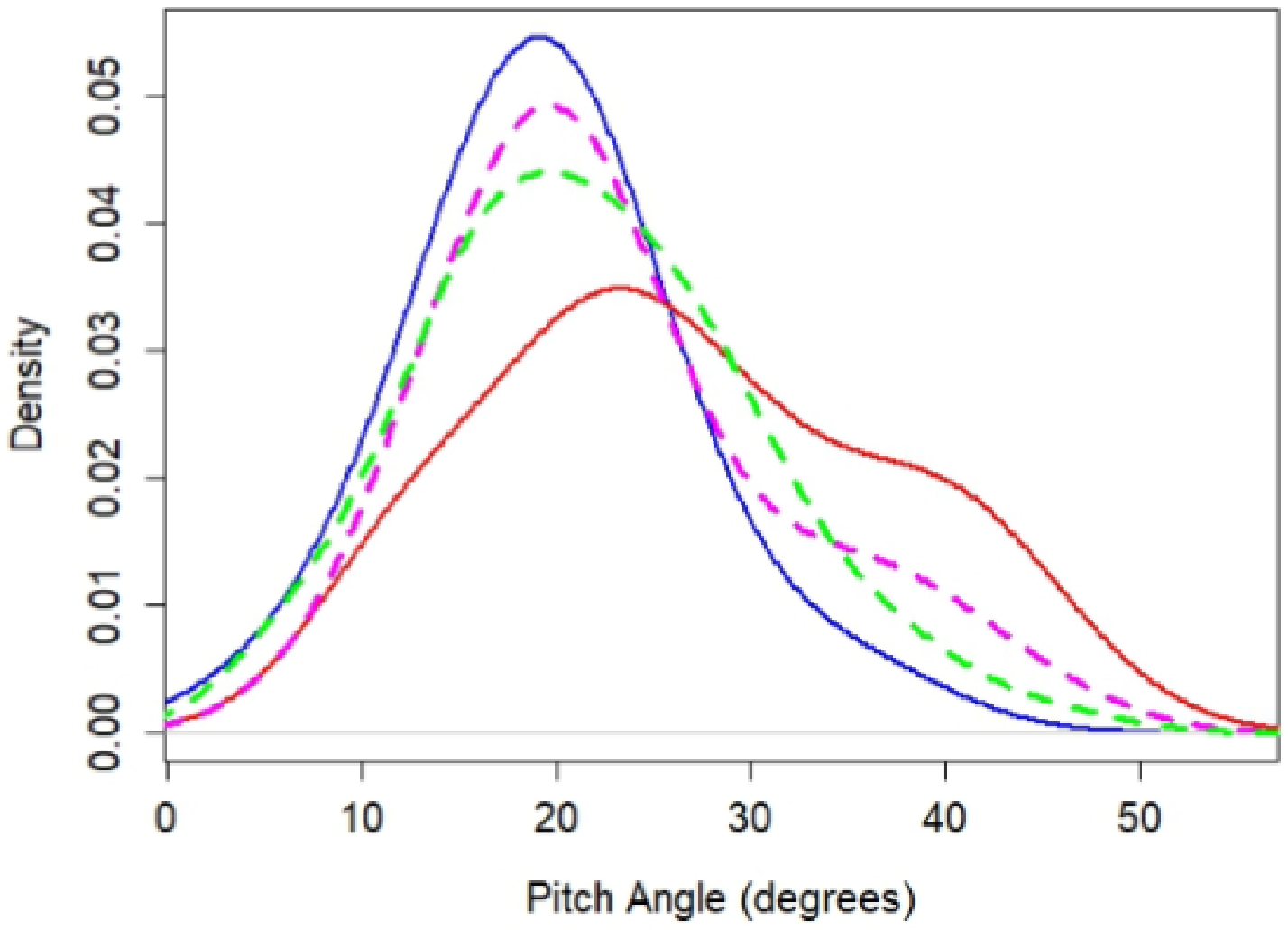}
     \includegraphics[width=9.18 cm]{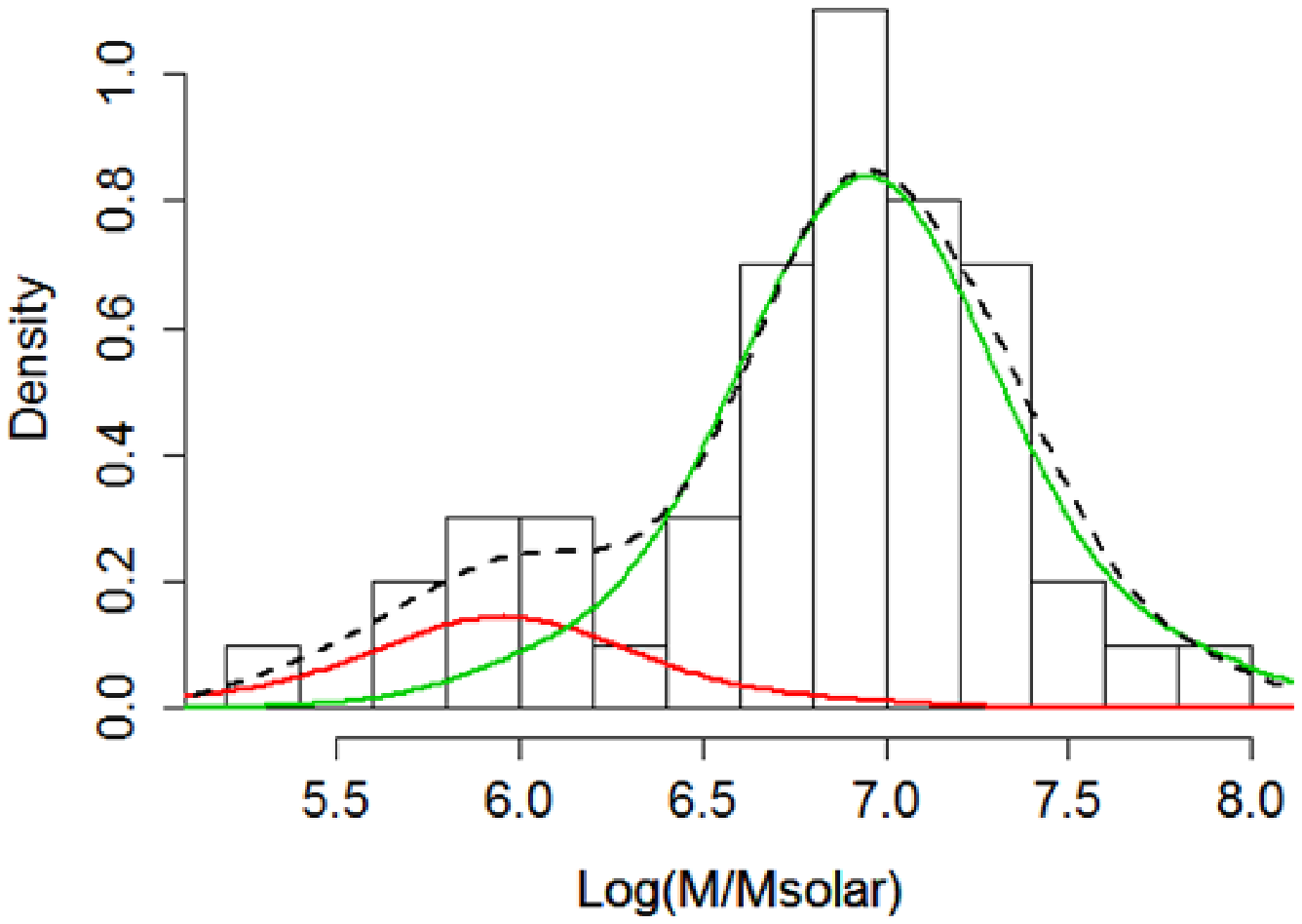}
     \end{adjustwidth}
		\caption{
		\textbf{(Left)} Corrected PADFs of key subsamples: bright sample (\textcolor{green}{{\hdashrule[0.35ex]{8mm}{1pt}{1mm}}}), faint sample (\textcolor{magenta}{{\hdashrule[0.35ex]{8mm}{1pt}{1mm}}}), faint subsample of Hubble types Sa--Sc (\textcolor{blue}{{\hdashrule[0.35ex]{8mm}{0.4mm}{}}}), and the faint subsample of Hubble types Scd \& Sd (\textcolor{red}{{\hdashrule[0.35ex]{8mm}{0.4mm}{}}}).
		From this figure it is clear that while the Scd \& Sd type galaxies make up a small fraction of the faint sample, they are the cause of most of the behavior at the tail of the distribution.
		It is also apparent that the excess peak height in the faint sample is a result of the comparative overabundance of Hubble type Sc galaxies, which cluster strongly around the peak.
		\textbf{(Right)} BHMF of the faint sample decomposed into low-mass (\textcolor{red}{{\hdashrule[0.35ex]{8mm}{0.4mm}{}}}) and high-mass (\textcolor{green}{{\hdashrule[0.35ex]{8mm}{0.4mm}{}}}) populations.
		Curves are semi-parametric fits to the cluster data.
		The dashed line (\hdashrule[0.35ex]{8mm}{1pt}{1mm}) is the KDE of original distribution.
		}
    \label{bhmf2}
\end{figure}

    To flesh out this relationship, we fit the KDE probability densities of the sample according to Hubble type (see Figure~\ref{bhmf2}, left).
    We find the Scd and Sd type galaxies indeed peak at higher pitch angles (lower SMBH masses).
    If the measured galaxies in the faint sample are divided by Hubble type, Sc types make up about half the sample with 25 measured galaxies, with Sa, Sab, Sb, and Sbc together making up about a quarter (all but one measured being Sb or Sbc) and Scd, Sd, and Sm types making up approximately the final quarter (no Sm galaxies were measured for the faint sample).
    When Sa, Sab, Sb, Sbc, and Sc types are grouped together with Scd, Sd, and Sm types are grouped separately, we find the tighter-pitch-angle, higher-mass group is well behaved with a very weak tail, but the Scd--Sd group has a broad peak shifted to higher pitch angles: $|\phi|(\max\rho)=23\fdg24$ and $\bar{|\phi|}=28\fdg69$, \textit{cf}. $|\phi|(\max\rho)=17\fdg50$ and $\bar{|\phi|}=22\fdg86$ for the entire dim sample.
    This is not unexpected, as Hubble types are defined only in part by the tightness of spiral arm winding, \textit{i.e.}, Hubble type letters past Sc should mostly correspond to loose spirals.\endnote{
    It is worth reviewing how the traditional Hubble-Jeans sequence \citep{Jeans:1919,Jeans:1928,Lundmark:1925,Hubble:1926,Hubble:1926b,Hubble:1927,Hubble:1936} was built primarily on the geometry of spiral winding \citep{Bergh:1998,Lapparent:2011}.
    However, modern spiral galaxy morphological classifications have become almost entirely based on central bulge size \citep{Graham:2008,Willett:2013,Masters:2019}.
    Yet, numerous studies have concluded that only a weak correlation exists between bulge-to-disk light ratio and Hubble type \citep{Courteau:1996,Jong:1996,Seigar:1998a,Patsis:2004}.
    Nonetheless, $|\phi|$ does \emph{decrease} monotonically with \emph{increasing} bulge-to-total light ratio \citep[\textit{e.g.},][Figure~5]{Yu:2019}.}

	Using the two clustering populations, we produce a possible two-population model for the probability density which matches closely with the global population (see Figure~\ref{bhmf2}, right).
	The faint sample population may be described with the superposition of two near-normal distributions or as one population with a heavy tail.
	In either case, the behavior at the extreme high-$|\phi|$/low-$M_{\bullet}$ end of the distribution is explained by the Scd and Sd type galaxies in the sample.
	Of course, additional data is needed to more definitively describe this apparent bimodal distribution.
	
\subsection{Combined BHMF for Local Spiral Galaxies, Accounting for Sample Completeness}

\begin{figure}[h]
	\includegraphics[clip=true, trim= 3mm 10mm 10mm 20mm, width=\columnwidth]{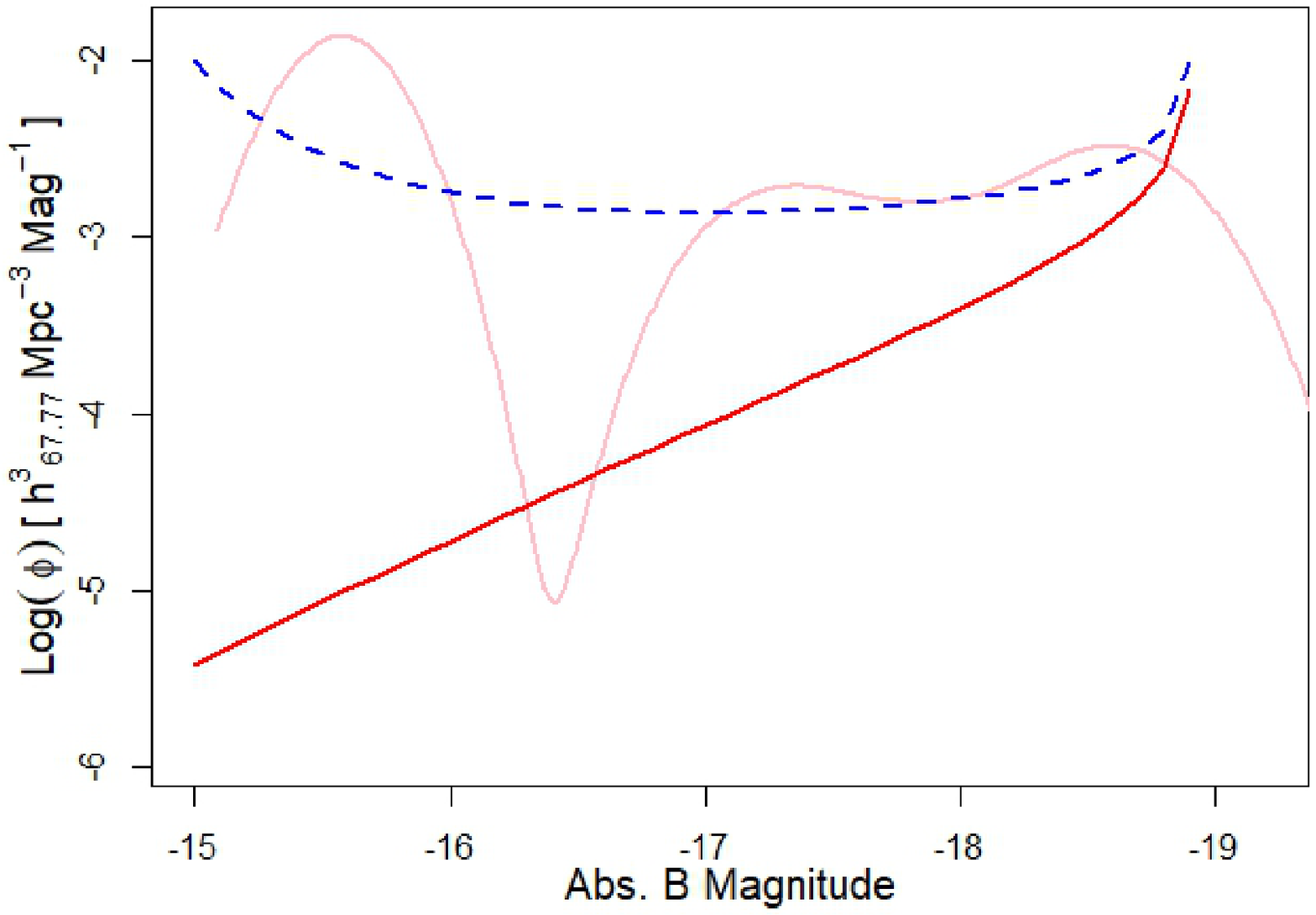}
		\caption{
		Luminosity functions for the \citetalias{2011ApJS..197...21H} faint galaxies: uncorrected luminosity function of the sample (\textcolor{red}{{\hdashrule[0.35ex]{7mm}{0.4mm}{}}}), luminosity function fit as corrected for the effective comoving volume of the galaxies in the sample (\textcolor{blue}{{\hdashrule[0.35ex]{8mm}{1pt}{1mm}}}), and KDE check of the corrected luminosity function, showing the probability density based on actual galaxies (\textcolor{pink}{{\hdashrule[0.35ex]{7mm}{0.4mm}{}}}).
		The number of measured galaxies at $\mathfrak{M}_{B}>-17$\,mag was very small, accounting for the spikes below and above the functional fit correction.
		The corrected luminosity function produces reasonable galaxy abundances for the corrected sample.
		Galaxies in this regime are weighted for density calculations with this correction to produce the BHMF for all measurable local spiral galaxies.
		}
    \label{lum}
\end{figure}

	In order to construct the BHMF of local spiral galaxies, we combine the bright sample from \citet{2014ApJ...789..124D} with the faint sample to eliminate the $\mathfrak{M}_{B}\leq-19.12$\,mag constraint on the mass function.
	To do this properly, galaxies missing from the faint sample due to detection limits must be accounted for.
	This is accomplished by weighting the galaxies from both samples based on the effective comoving volume of galaxies by $\mathfrak{M}_{B}$.
	The fainter the galaxies targeted, the smaller effective comoving volume in which such a galaxy could be detected.
	Our corrected luminosity function uses a Pearson distribution fit of the number of galaxies by magnitude with the effective comoving volume to up-correct the number of galaxies at very faint magnitudes (Figure~\ref{lum}).
	Thus, the luminosity function we present in Figure~\ref{lum} is complementary to the luminosity function in \citet[][Figure~2]{2014ApJ...789..124D}; Figure~\ref{lum} shows the number density for $\mathfrak{M}_{B}>-19.12$\,mag, whereas \citet[][Figure~2]{2014ApJ...789..124D} shows the number density for $\mathfrak{M}_{B}\leq-19.12$\,mag.
	We check the Pearson distribution fit against a KDE and find a reasonable match to the data, providing confidence in the computed luminosity function.
	
\begin{figure}
	\includegraphics[clip=true, trim= 0mm 5mm 10mm 20mm, width=\columnwidth]{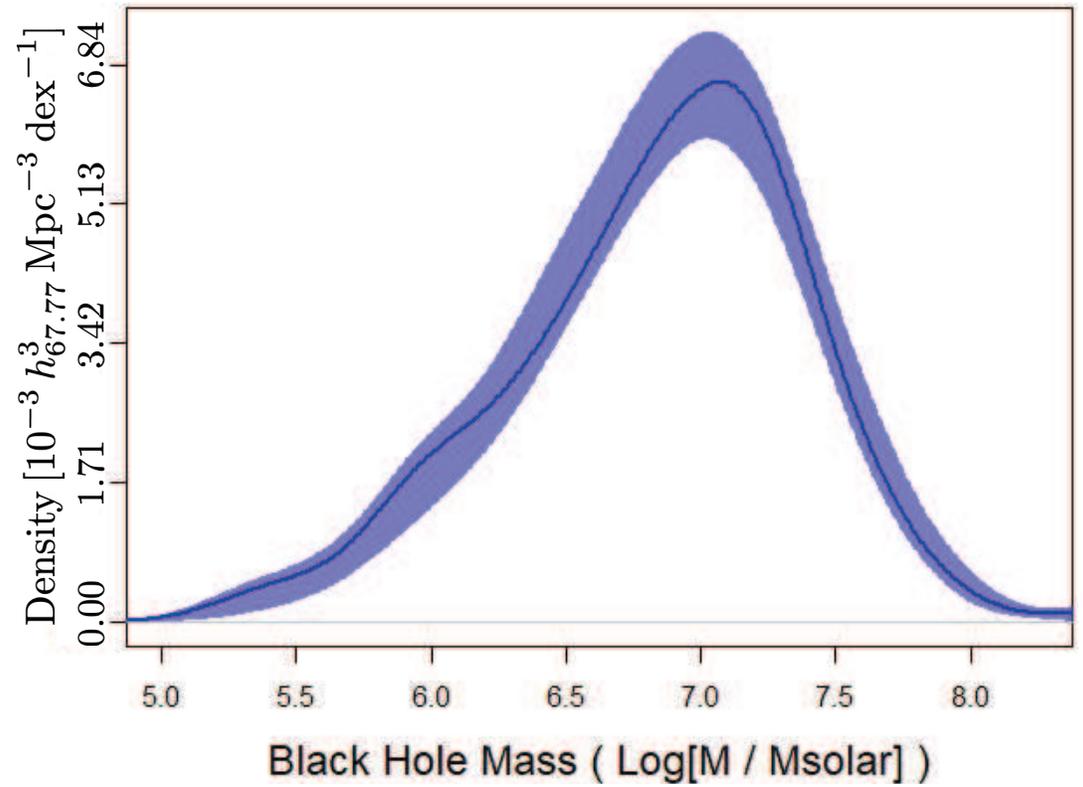}
		\caption{
		Combined BHMF for spiral galaxies, with 90\% confidence intervals and with a limiting luminosity distance of 25.4\,Mpc.
		With the addition of the complementary sample from \citet{2014ApJ...789..124D}, this BHMF is representative of spiral galaxies of all Hubble types.
		}
    \label{mco}
\end{figure}

    With this new weighting based on effective comoving volume, it is possible to properly combine the complete sample \citep{2014ApJ...789..124D} and the adjusted incomplete sample into one population which describes the BHMF for spiral galaxies in the local Universe (Figure~\ref{mco}).
    The BHMF is calculated in the same manner as with each individual sample, but with a weighting of the combined input data to account for missing galaxies.
    The result of this combination retains some of the properties for each of the samples from which it originated, as it is a convolution of the two distributions.
    As such, the peak location remains largely unchanged.
    For comparison, the density at $M_{\bullet}=10^{7.00}$\,M$_{\sun}$ for the bright sample was presented as $2.82^{+0.22}_{-0.26}\times10^{-3}\,h^3_{67.77}$\,Mpc$^{-3}$\,dex$^{-1}$ in \citet{2014ApJ...789..124D}, and we now increase that density to $6.59^{+0.40}_{-0.39}\times10^{-3}\,h^3_{67.77}$ by adding in and accounting for the population of dimmer galaxies.
    There is some retention of a heavier tail to the distribution at high-$|\phi|$/low-$M_{\bullet}$.
    The combined BHMF fully-describes the distribution of black holes in spiral galaxies in the local Universe.
    
\section{Discussion}\label{sec:discussion}
	
	\subsection{A Bimodal Distribution}\label{sec:bimodal}
	
	The PADF and BHMF established from our latest sample of local disk galaxies resembles those in \citet{2014ApJ...789..124D}.
	These subsets are mirrored at low-$|\phi|$/high-$M_{\bullet}$.
	However, they may diverge at the high-$|\phi|$ end of the dataset.
	The faint sample has a higher probability of large pitch angles (loose spirals) and a reduced probability of containing small pitch angles (tight spirals).
	The difference between the two samples as denoted by our KDE approach is small, but statistically significant (see Figure~\ref{bhmf2}).
	Insomuch, the morphological differences between the galaxies on the heavy tail of the distribution and those at the peak do suggest two separate populations.
	
	Indeed, \citet[][Figure~20]{Yu:2020} also find a distinct bimodal distribution in the PADF.
	Drawing a large sample of pitch angle measurements for 3569 spiral galaxies, they find the overall PADF exhibits $\bar{|\phi|}=18\fdg8\pm6\fdg7$, but is best described by the sum of two normal distributions: $\bar{|\phi|}=12\degr\pm3\fdg4$ and $\bar{|\phi|}=23\degr\pm4\fdg3$.
	Notably, the latter distribution is similar to the mean of our entire dim sample, $\bar{|\phi|}=22\fdg86$.
	However, the \citet{Yu:2020} distribution approximates the range of galaxies earlier than Hubble subtype Sd and has limited statistics at $\lesssim$$10^{9.5}$\,M$_{\sun}$ in galaxy stellar mass.
	Thus, they do not perceive the later population of galaxies around $\bar{|\phi|}=28\fdg69$ in the heavy tail of the distribution as we indicate.
	
	The \citetalias{2011ApJS..197...21H} sample contains a detectable difference in Hubble types across $\mathfrak{M}_{B}$, which illustrates the major difference between the two samples in question.
	The faint sample is largely missing its Sa and Sab types, and as a result has an excess of later types.
	The \citet{2014ApJ...789..124D} sample, meanwhile, is missing the Sm types and perhaps some Scd and Sd types as well, given the luminosity limit.
	The excesses and scarcities of the samples are borne out in density estimations of their PADF and BHMF.
	
	In comparison to the \citet{2014ApJ...789..124D} sample, the faint sample expresses scarcity at low $|\phi|$ (high $M_{\bullet}$) and overabundance at the distribution's peak, with hints of slight overabundance at the distribution's tail (\textit{i.e.}, too many \emph{average} spirals).
	The tail of the faint sample is heavy, suggesting the possibility of a multimodal distribution.
	We show that the Scd and Sd galaxies cluster apart from the Sa and Sab types in $\mathfrak{M}_{B}$, and largely comprise the high-$|\phi|$ subgroup found by bimodal clustering (see Figure~\ref{bhmf2}).
	Removal of this subgroup and replacement with typical Sa and Sab galaxies from the faint sample would likely cause the faint sample and the \citet{2014ApJ...789..124D} sample to have convergent BHMFs.
	
	As a whole, the resulting BHMFs of our sample and \citet{2014ApJ...789..124D} can be construed to be nearly identical once differences in population subgroups are accounted for.
	The weight of the low-mass tail of the BHMF remains somewhat unknown.
	Due to missing faint galaxies and poor pitch angle measurement rates in this regime, we are likely underestimating the density of the BHMF at its tail.
	The pitch angle measurement success rate for Sa--Sc types was better than the Scd--Sm type galaxies.
	Since the distribution for Scd and Sd types peak at higher $|\phi|$, it follows that the tail of the distribution may be heavier than we measure.
	
	By correcting our sample for effective comoving volume, we attempt to recover a more representative distribution of galaxy pitch angles by accounting for galaxies missed due to their low surface brightness.
	In so doing, a combined sample is produced containing all galaxies within the \citetalias{2011ApJS..197...21H} sample whose pitch angles were measured and whose limiting luminosity distance was 25.4\,Mpc.
	This combined sample is representative of the local BHMF of spiral galaxies in the local Universe.
	However, as aforementioned, our estimate of the relative density at high-$|\phi|$/low-$M_{\bullet}$ represents an improvement on previous work, but should be considered a lower limit.
	Verily, there is still much to learn yet about the true nature of the ``left'' side of the BHMF.
	Indeed, much uncertainty persists at the low-mass end of the BHMF, as contemporary observational \citep[\textit{e.g.},][]{Gallo:2019} and theoretical \citep[\textit{e.g.},][]{Habouzit:2021} studies of the BHMF both continue to be hampered in finding a consensus shape (either rising or falling) at the low-mass end.
	
	\subsubsection{The Impact of the Later Hubble Type Spiral Galaxies}
	
	It is understood that the \emph{right} side of the Hubble tuning fork is reserved for the \emph{latest} morphologies of spiral galaxies, \textit{i.e.}, galaxies with small bulges (or bulgeless) and the most open winding of spiral arms (high $|\phi|$).
	This structure is also represented in the modern galaxy classification grid of \citet{Graham:2019b}.
	 The prevailing theory behind the genesis of spiral structure is described by density waves, the propagation of which is determined by the bulge mass, the disk density, and the spiral-arm pitch angle \citep{Lin:1966,Davis:2015}.
	 It is also well established the that mass of black holes are correlated with the mass of their host bulges \citep[\textit{e.g.},][]{2019ApJ...873...85D,Sahu:2019a}.
	 Ergo, it is expected the smallest central black holes will be hosted by these late-type galaxies with the highest morphological stages.
	 Thus, a key result of our study is that indeed, the low-end of the BHMF ($M_{\bullet}\lesssim10^6\,\mathrm{M_{\sun}}$) is dominated by the marginal population of these galaxies with high Hubble stages (see Figure~\ref{bhmf2}, bottom panel).
		
	The complete magnitude- and volume-limited subsample ($\mathfrak{M}_{B}\leq-19.12$\,mag and $D_{\mathrm{L}} \leq 25.4$\,Mpc) of the \citetalias{2011ApJS..197...21H} consists of 68 early-type (30 elliptical $+$ 38 lenticular) galaxies \citep{Mutlu-Pakdil:2016} and 140 late-type (\textit{i.e.}, spiral) galaxies \citep{2014ApJ...789..124D}, for a grand total of 208 galaxies.
	Of the 208 galaxies, there are five Scd, one Sd, and three Sm galaxy types.
	Thus, only 9/208 (4.33\%) galaxies in the local Universe are Scd types or later ($T\geq6$).\endnote{
	For comparison, the full \citetalias{2011ApJS..197...21H} sample (southern hemisphere galaxies with $B_{\rm T}<12.9$\,mag) plus the Milky Way has 41/606 (6.77\%) galaxies with $T\geq6$.}
	However, \citet{Lacerda:2020}'s diameter-limited sample of galaxies shows that 163/867 (18.8\%) galaxies are $T\geq6$.
	
	Moreover, the much larger diameter-limited sample of the Third Reference Catalogue of Bright Galaxies \citep[][hereafter \citetalias{RC3}]{RC3} has 5,480/17,801 (30.8\%) galaxies with $T\geq6$.\endnote{
	The EFIGI catalogue \citep{Baillard:2011} of 4,458 galaxies is a subset of the \citetalias{RC3}, but no longer considers peculiar galaxies or galaxies with special features as belonging to separate stages.
	As a result, they classify an even higher fraction (1,665/4,45$8\approx3$7.3\%) of galaxies as $T\geq6$ \citep{Lapparent:2011}.}
	This observed growing number of less massive galaxies is also reflected in the galaxy stellar mass function.
	\citet{Driver:2022} show that the total number density of local galaxies ($z<0.1$) monotonically \emph{increases} as the stellar mass of galaxies \emph{decreases} (measured as low as $10^{6.75}$\,M$_{\sun}$\,$h_{70}^{-2}$), as does the number densities of the morphological subsamples of ``diffuse-bulge plus disk'' and ``disk-only'' galaxies ($z<0.08$).
	Therefore, the growing fraction of later type galaxies with increasing sample size indicates that the true number of later type galaxies is quite large and our value of 4.33\% is, by design, a lower limit.
	
	Importantly, opposite trends are observed between the galaxy stellar mass function and the galaxy stellar mass density function (and for the aforementioned morphological subsamples).
	While the number density of galaxies continuously \emph{increases} as galaxies become \emph{less} massive, the mass density of galaxies continuously \emph{decreases} as galaxies become progressively \emph{less} massive than the peak of the distribution at $10^{10.8}$\,M$_{\sun}$\,$h_{70}^{-2}$ \citep[see][Figure~12]{Driver:2022}.
	As \citet{Driver:2022} point out, this implies that low-mass galaxies are the most numerous in our local Universe, but the product of their low mass with high number density yields a minimal contribution to the total stellar mass density.
	Because low-mass black holes are expected to reside in low-mass galaxies, this same opposing trend between the low-mass ends of the BHMF and the black hole mass density function is also expected, but not (yet) observed.
	That is, \citet{2014ApJ...789..124D} find that both the BHMF and the black hole mass density function decline at the low-mass end.
	Thus, this represents an opportunity to search for the unobserved population of low-mass black holes in our local Universe.
	
	One sure way of knowing a black hole is present at the center of a galaxy is by detecting an active galactic nucleus (AGN), which is indicative of an accreting black hole.
	Considering \citet{Lacerda:2020}'s 163 ($T\geq6$) galaxies, only 1/163 (0.61\%) have an AGN.
	In contrast, 33/704 (4.69\%) of their $T<6$ galaxies host an AGN.
	A BHMF constructed using only 34 total AGNs from their sample would thus, only contribute at a proportion of 1/34 (2.94\%) to the low-mass end of the BHMF ($M_{\bullet}\lesssim10^6\,\mathrm{M_{\sun}}$).
	Therefore, it is these small percentages of later type galaxies that most influence the low-mass end of the BHMF, where the missing population of intermediate-mass black holes is hiding.
	
	\subsection{Intermediate-mass Black Holes}
	
Astronomers have discovered two sizes of black holes in the Universe: SMBHs ($\geq$10$^5$\,M$_{\sun}$) and solar-mass black holes ($<$10$^2$\,M$_{\sun}$).
However, the so-called intermediate-mass black holes (IMBHs) theorized to fill in the evolutionary gap ($10^2\,\mathrm{M_{\sun}}\leq M_{\bullet}<10^5\,\mathrm{M_{\sun}}$), remain elusive.
The further discovery of IMBHs would be a great boon to research/theory and provide potential multi-messenger targets for future generations of gravitational wave detectors \citep[\textit{e.g.},][]{LISA:2022}, which will be sensitive to black holes in the intermediate-mass range \citep{Klein:2016}.
As such, the use of black hole mass scaling relations like the $M_{\bullet}$--$\phi$ relation can prove useful in helping to identify IMBH candidates in galaxies.
For an added benefit, as the number of galaxies with directly-measured IMBHs increases, so will the accuracy of a continually-refined $M_{\bullet}$--$\phi$ relation as its range of interpolation extends into the IMBH regime.

\citet{Koliopanos:2017} demonstrated the benefit of utilizing multiple scaling relations, including X-ray and radio luminosity relations, to search for possible IMBHs.
Subsequent research has included the \textit{Chandra} Virgo Cluster Survey of Spiral Galaxies \citep{Soria:2022}, which presents an analysis of the ultraluminous X-ray source population in 75 Virgo Cluster late-type galaxies, including all those with a star formation rate $\gtrsim$1\,M$_{\sun}$\,yr$^{-1}$.
\citet{Graham:2019,Graham:2021b} predicted the central black hole masses in these galaxies using the latest black hole mass scaling relations involving $\phi$, $\sigma$ \citep{Sahu:2019}, and $M_{\rm \star,tot}$ \citep{Davis:2018,Sahu:2019a}.
\citet{Graham:2019} identified three late-type galaxies in the Virgo Cluster with both a predicted black hole mass of $\lesssim$10$^5$\,M$_{\sun}$ and a centrally located X-ray point source.
Subsequently, \citet{Graham:2021b} revealed 11 more such galaxies, more than tripling the number of active IMBH candidates among late-type galaxies of the Virgo Cluster.\endnote{
Additionally, \citet{Graham:2021a} presented a serendipitous discovery of a potential shredded offset nuclear star cluster with an IMBH, identified by an X-ray source and optical/infrared counterpart.}
Moreover, \citet{Davis:2021} demonstrated a novel method of combining ten different black hole mass estimates for a single galaxy (NGC~3319) in order to produce a probability density function of black hole mass for the galaxy, and thus a refined mass estimate with a better precision than that of individual estimates.
Ongoing research \citep{Davis:2022} will repeat this procedure and measure pitch angles for all (85) of the later type spiral galaxies (\textit{i.e.}, Hubble type Sd) in the \citetalias{RC3}, in a further search for IMBH candidates.
    
    The low-mass, late-type spiral galaxy portion of the population studied here represents a strong candidate set for further study toward characterizing IMBHs.
    As such, pitch angle measurements can easily identify the late-type spirals with very loose (high-$|\phi|$) arms, indicating good candidates for further IMBH study.
    Specifically, \citet{Davis:2017} find that IMBHs are mostly likely to reside in spiral galaxies with $|\phi|\gtrsim26\fdg7$,\endnote{
    See \citet[][Figure~1]{Davis:2021}, for a graphical representation of a logarithmic spiral with $|\phi|=26\fdg7$.} corresponding to masses $M_{\bullet} \lesssim10^5\,{\rm M}_{\sun}$ and likely fall into the IMBH range.
    However, given that we find pitch angles become more difficult to measure as $|\phi|$ increases (see Figure~\ref{heterosca}), we advocate utilizing the $M_{\bullet}$--$\phi$ relation in tandem with other black hole mass scaling relations to more precisely estimate masses of potential IMBHs \citep[\textit{e.g.},][]{Davis:2021}.
    
    Based upon our conservative analysis, we find six galaxies from our sample (IC~2627, IC~5201, NGC~247, NGC~1487, NGC~1518, and NGC~4592; see Table~\ref{table}) with potential IMBHs ($M_{\bullet}-\delta M_{\bullet}<10^5\,{\rm M}_{\sun}$).
    Specifically, we find that NGC~1487 is our strongest candidate for hosting an IMBH, with an estimated black hole mass of $8.0_{-\;\,6.6}^{+38.3}\times10^3\,{\rm M}_{\sun}$.
    At the present time, the dwarf elliptical galaxy NGC~205 (M110), which is a satellite of the Andromeda Galaxy (M31), has the least massive
nuclear black hole as measured via direct methods.\endnote{
This clear example with NGC~205 draws attention to dwarf early-type galaxies as good candidates to host IMBHs.
Furthermore, dwarf early-type galaxies occasionally possess faint disk substructure, including bars and spiral arms, and thus, can yield pitch angle measurements \citep{Jerjen:2000,Lisker:2006,Michea:2021}.
However, their typically elusive spiral structure often requires significant image processing to extract any extant embedded disk component.
Specifically, 41/476 ($\approx$$8.6\%$) dwarf early-type galaxies in the sample of \citet{Lisker:2006} have ``possible, probable, or unambiguous disk features.''
Although, as \citet{SmithR:2021} point out, volute structure in dwarf early-type galaxies could be generated by tidal triggering as the product of the cluster harassment of passive dwarf galaxies.
}
    Thus, we find our estimated mass for a central black hole in NGC~1487 is on par with that found in NGC~205 via stellar dynamical modeling, $M_{\bullet}=6.8_{-\;\,2.2}^{+31.9}\times10^3\,{\rm M}_{\sun}$ \citep{2019ApJ...872..104N}.
    Together, these aforementioned half-dozen galaxies will be the subject of further investigation with varied black hole mass scaling relations and to see if they possess additional indicators of an actively accreting black hole at the centers (\textit{e.g.}, X-ray point source detection).
    
    \subsection{Spiral Pattern Formation}
    
Besides the creation of spiral patterns in galaxies via tidal interactions \citep{Toomre:1972,Tully:1974,Elmegreen:1983,Oh:2008,Dobbs:2010}, there are two main competing theories for the genesis of spirals: (i) quasi-stationary density waves \citep[][for a review]{Lindblad:1963,1964ApJ...140..646L,2016ARA&A..54..667S} and (ii) local instabilities/perturbations that develop into dynamic material arms via ``swing amplification'' \citep[][for a review]{Goldreich:1965,Julian:1966,Goldreich:1978,Toomre:1981,Sellwood:1988,Dobbs:2014}.
As mentioned in our \S\ref{sec:bimodal}, \citet{Yu:2020} find a clear bimodal distribution in their PADF that roughly agrees with our results.
\citet{Yu:2020} reflect on the possibility that their later-type population of spiral galaxies ($\bar{|\phi|}=23\degr\pm4\fdg3$) could be the result of swing amplification.
They draw on the models of \citet{Hart:2018} that produce a population of galaxies built by swing amplification with a median $|\phi|=24\degr$.
Similarly, our dim sample of spiral galaxies could at least be partially governed by swing amplification.
That is, \citet{Hart:2018} find that 60\% of the galaxies in their sample \emph{cannot} be explained by the expected characteristics of swing amplification, and more likely, a multitude of mechanisms operate to trigger and moderate spiral arms in the majority of spiral galaxies.
Therefore, our dim sample could be more influenced by the tenets of swing amplification than its bright counterpart \citep{2014ApJ...789..124D}, and likely leads to a greater scatter resultant from the application of scaling relations.
    
Building upon the observational study of \citet{Davis:2015}, further studies have cemented spiral density waves as the most probable mechanism for the genesis of spiral patterns in disk galaxies.
\citet[][\textit{cf.} \citealt{Yu:2018b}]{Hamed:2016} demonstrated that a perceptible difference exists between pitch angle measurements across the electromagnetic spectrum; $|\phi|$ is systematically higher in images that highlight star formation (\textit{e.g.}, far-infrared and ultraviolet) and lower in images that highlight aging stellar populations (\textit{e.g.}, near-infrared).\endnote{
For consistency, all pitch angle measurements in our study were made from optical $B$-band images.}
\citet{Miller:2019} go further and show that indeed, the $|\phi|$ spectral differences observed by \citet{Hamed:2016} are caused by enhanced regions of stellar light downstream from star-forming regions, as predicted by density wave theory.
In turn, these results are further tested and affirmed by subsequent studies of co-rotation radii \citep{Abdeen:2020} and age gradients \citep{Abdeen:2022}.
Furthermore, \citet{Smith:2022} present relations between pitch angle, concentration, and specific star formation rate, which they say supports the fundamental plane suggested by \citet{Davis:2015} by assuming central concentration increases with bulge mass and since star formation rate increases with increasing gas surface density \citep{Schmidt:1959,Kennicutt:1989,Kennicutt:1998}.
    
\citet{Sellwood:2014,2019MNRAS.489..116S} provide evidence of self-excited instabilities as the cause of spiral patterns in simulations of unperturbed disks.\endnote{
In contrast to the cases of unperturbed disks, the simulations by \citet{Kumar:2022} demonstrate that fly-by interactions excite strong spirals in the outer regions of a host galaxy's disk, but the spirals tighten after pericenter passage of a perturbing galaxy and fade away after a few Gyr.
}
They find that although individual spiral modes are transient \citep{Sellwood:2011}, fresh instabilities are excited as the old modes fade away, resulting in a recurrent instability cycle.
If spirals are primarily driven by density waves, the newly-birthed modes should have the same pitch angle as their dying parents, which has been demonstrated in simulations \citep{Jazmin:2013,Jazmin:2014,Jazmin:2015,Jazmin:2015b}.
If this is the case, pitch angle measurement as a method for probing other galaxy characteristics is likely to be robust for a given galaxy across time, perhaps with the exception of transitional periods between dominant modes.

As previously mentioned, there exists a planar relationship among the quantities of central bulge stellar mass, pitch angle, and neutral hydrogen gas density in the galactic disk \citep{Davis:2015}.
It has also been shown that higher disk masses favor lower multiplicity spiral patterns \citep{1984ApJ...282...61S,1987A&A...179...23A,2015ApJ...799..213B}.\endnote{For a contrasting view, see also \citet{Hart:2017b} and \citet{Porter-Temple:2022}, who find diminished star formation rates in many-armed spiral galaxies.}
Galaxies with lower disk densities should, then, have a tendency toward a higher number of arms and exhibit less $m=2$ grand design spirals.
Indeed, \citet{Yu:2020} empirically show (see their Figure~17a) that $m\propto T$ (albeit with considerable scatter), \textit{i.e.}, the number of spiral arms increases monotonically as the Hubble stage increases.
However, lower disk density galaxies \emph{could} produce spirals with similar pitch angles to their heavy disk counterparts \emph{if} their bulges are less massive.

Going forward, a census of neutral hydrogen gas density in galaxies targeted for pitch angle measurement should be taken, such that the populational differences among these parameters (and their effects) may be better explored.
This is especially relevant for galaxies without grand design spiral structure, as their disks are likely to be less massive.
As \citet{Hart:2017} point out, \emph{the central mass concentration of a galaxy may not be the dominant cause of pitch angle}, which is more likely to be important for galaxies with small bulges.
Specifically, \citet{Masters:2019} find, ``Galaxies with small bulges have a wide range of arm winding, while those with larger bulges favor tighter arms.''
Yet the general trend instilled by the Hubble sequence still holds (albeit with considerable scatter), \textit{i.e.}, $|\phi|\propto T$ \citep{Kennicutt:1981,Seigar:1998,Yu:2018,Garcia:2019,Yu:2020}.\endnote{
\citet{Treuthardt:2012} demonstrate that the correlation between $|\phi|$ and $T$ is tightest for spiral galaxies with fast rotating bars.
Moreover, the tight relation found by \citet{Treuthardt:2012} is in close agreement with the theoretical relation between $|\phi|$ and $T$ presented by \citet{Roberts:1975}.
}
In order to accurately produce the BHMF for late-type galaxies, an alteration of pitch angle due to different gas densities should be taken into account.
Until then, unaccounted disk densities likely manifest by contributing to the uncertainty and scatter in the $M_{\bullet}$--$\phi$ relation, and subsequently its use to generate a BHMF.

\section{Conclusions}\label{sec:Conclusions}
	
Our latest sample of \citetalias{2011ApJS..197...21H} spiral galaxies has a PADF (and corresponding BHMF) analogous to that of \citet{2014ApJ...789..124D}.
As expected, there is strong indication of high-mass black holes missing from the faint sample, and some indication of an excess of lower mass black holes on the distribution's tail.
It may be claimed, although populational differences exist between the two samples, the BHMF of spiral galaxies in the local Universe is not strongly altered at its low-mass end by the inclusion of galaxies with Scd and Sd Hubble types.
However, if we consider our additions to the BHMF \citep{2014ApJ...789..124D} and the galaxy stellar mass function \citep{Driver:2022} as guides, it leads us to the conclusion that many low-mass black holes (\textit{i.e.}, IMBHs) in our local Universe remain undiscovered. 
Excesses at the low-mass tail of the BHMF may be larger than recorded in our data, due to galaxies too faint to be observed and unmeasurable pitch angles for the latest Hubble types.
Following on the claim that galaxies with $|\phi|\gtrsim26\fdg7$ are candidates for hosting IMBHs \citep{Davis:2017}, it may be the case that the heavy tail of our pitch angle distribution is composed, in part, of galaxies hosting these objects.

We would like to highlight the following key results presented in this paper:
\begin{itemize}
\item We confirm the existence of a bimodal distribution in the PADF.
\item The PADF (and resulting BHMF) for late-type galaxies is dominated by the intrinsic minority of $T<6$ galaxies, but mostly overlooks the intrinsic majority of $T\geq6$ galaxies that constitutes the second population in the bimodal distribution.
\item This second population of galaxies ($T\geq6$) is of particular interest for its preponderance of potential IMBHs.
\item Finally, the morphological demarcation between the two populations of spiral galaxies could be indicative of competing spiral formation mechanisms.
\end{itemize}

Our work here provides compelling evidence of a ``silent majority'' of low-mass galaxies, with their low-mass black holes, that are not adequately represented in the BHMF.
While mainly qualitative in nature at this time, our study calls for continued investigation and quantitative accounting of the true population of ``less-than-supermassive'' black holes ($M_{\bullet}\lesssim10^6\,\mathrm{M_{\sun}}$).
Additionally, the performance of pitch angle measurement techniques as a function of galaxy morphology, luminosity, and redshift, is yet to be surveyed.
Studies in these areas, as well as in selection effects and biases, should reduce measurement errors and improve measurement success rates for future work.
For now, we largely reaffirm the BHMF for local spiral galaxies \citep{2014ApJ...789..124D}, with the possibility of another population of lower mass black holes (including IMBHs) amongst the very late Hubble types and exhibited in the high-$|\phi|$ tail of the distribution.



\vspace{6pt} 



\authorcontributions{
The authors’ respective contributions are as follows: conceptualization, M.S.F. and J.K.; methodology, M.S.F., J.K, and B.L.D.; software, M.S.S. and B.L.D.; validation, all authors; formal analysis and investigation, M.S.F.; resources, M.S.S.; data curation, M.S.F. and B.L.D.; writing---original draft preparation, M.S.F.; writing---review and editing, all authors; visualization, M.S.F.; supervision, J.K.; project administration, J.K.; funding acquisition, J.K. and B.L.D.
All authors have read and agreed to the published version of the manuscript.
}


\funding{
This research was funded by a grant from the Arkansas Space Grant Consortium. 
The Australian Research Council's funding scheme DP17012923 supported this research.
Parts of this research were conducted by the Australian Research Council Centre of Excellence for Gravitational Wave Discovery (OzGrav), through project number CE170100004.
This material is based upon work supported by, and the APC was funded by, Tamkeen under the NYU Abu Dhabi Research Institute grant CAP$^3$.
}



\institutionalreview{Not applicable.}



\informedconsent{Not applicable.}


\dataavailability{
Our data sample and basic properties are derived from the \citetalias{2011ApJS..197...21H}.
This research has made use of the NASA/IPAC Extragalactic Database (NED), which is funded by the National Aeronautics and Space Administration and operated by the California Institute of Technology.
The data products presented in this study are available on request from the corresponding author.
Note that unfortunately, this project suffered a significant loss of ancillary data.
As a result, all of the metadata related to the pitch angle measurements (\textit{e.g.}, the number of spiral arms, deprojection parameters, pitch angles as a function of radius, auxiliary post-processing files and records, \textit{etc.}), and reported throughout this paper in aggregate, is now lost for all of the galaxies in this study.
However, these metadata do not impact the main topic of this paper.
}


\acknowledgments{
B.L.D. thanks David Nelson for the use of his secluded office space during the COVID-19 pandemic.
This research made use of
\textsc{2dfft}~\citep{2012ApJS..199...33D,2016ascl.soft08015D},
\textsc{2dfft Utilities}~(\url{https://github.com/AGES-UARK/2dfft_utils}),
\textsc{2dfftUtils Module}~(\url{https://github.com/ebmonson/2DFFTUtils-Module}),
\textsc{daophot}~\citep{1987PASP...99..191S},
\textsc{fitdistrplus}~\citep{R2},
\citet{R4},
\textsc{galfit}~\citep{2010AJ....139.2097P},
\textsc{ggplot2}~\citep{R3},
\textsc{iraf}~\citep{Tody:1986,IRAF},
\textsc{ellipse}~\citep{1987MNRAS.226..747J},
\textsc{kedd}~\citep{R7},
the NASA/IPAC Extragalactic Database (NED),
the NASA Astrophysics Data System (ADS),
\textsc{pearson\_ds}~\citep{R6},
\textsc{pearspdf}~(\url{https://www.mathworks.com/matlabcentral/fileexchange/26516-pearspdf}),
\textsc{r}~\citep{R},
\textsc{SAOImageDS9}~\citep{Joye:2003}, and
\textsc{SpiralArmCount}~\citep{Spirality,Shields:2015}.
}


\conflictsofinterest{
The authors declare no conflict of interest.
The funders had no role in the design of the study; in the collection, analyses, or interpretation of data; in the writing of the manuscript; or in the decision to publish the~results.
}



\abbreviations{Abbreviations, Acronyms, Initialisms, and Truncations}{
The following abbreviations, acronyms, initialisms, and truncations are used in this manuscript:\\

\noindent 
\begin{tabular}{@{}ll}
\textsc{2dfft} & Two-dimensional Fast Fourier Transform\\
A\&A & Astronomy \& Astrophysics\\
AC & Arm Class\\
ADS & Astrophysics Data System\\
AGES & Arkansas Galaxy Evolution Survey\\
AGN & Active Galactic Nucleus\\
AHPCC & Arkansas High Performance Computing Center\\
AJ & Astronomical Journal\\
\textit{AL} & \textit{Alii}\\
APC & Article Processing Charge\\
APJ & Astrophysical Journal\\
APJS & Astrophysical Journal Supplement Series\\
AR & Arkansas\\
ARA\&A & Annual Review of Astronomy and Astrophysics\\
ASCL & Astrophysics Source Code Library\\
ASTRON & Astronomical\\
BHMF & Black Hole Mass Function\\
CA & California\\
CALIFA & Calar Alto Legacy Integral Field Area Survey\\
CAP$^3$ & Center Astro, Particle, and Planetary Physics\\
CC & Creative Commons\\
CCD & Charge-coupled Device\\
\textit{CF} & \textit{Confer}\\
CGS & Carnegie-Irving Galaxy Survey\\
COM & Commercial\\
COVID-19 & Coronavirus Disease 2019\\
CT & Connecticut\\
CV & Cataclysmic Variables\\
DOI & Digital Object Identifier
\end{tabular}

\noindent 
\begin{tabular}{@{}ll}
ED & Edition\\
EDS & Editions\\
EDU & Educause\\
EFIGI & Extraction de Formes Idéalisées de Galaxies en Imagerie\\
\textit{EG} & \textit{Exempli Gratia}\\
EM & Expectation-maximization\\
ENG & English\\
ESO & European Southern Observatory\\
\textit{ETC} & \textit{Et Cetera}\\
GAMA & Galaxy and Mass Assembly\\
HTTPS & Hypertext Transfer Protocol Secure\\
IC & Index Catalogue of Nebulae and Clusters of Stars\\
\textit{IE} & \textit{Id Est}\\
IMBH & Intermediate-mass Black Hole\\
IPAC & Infrared Processing and Analysis Center\\
IR & Infrared\\
\textsc{iraf} & Image Reduction and Analysis Facility\\
KDE & Kernel Density Estimator\\
M & Messier\\
MAX & Maximum\\
MDPI & Multidisciplinary Digital Publishing Institute\\
MIN & Minimum\\
MLCV & Maximum Likelihood Cross Validation\\
MNRAS & Monthly Notices of the Royal Astronomical Society\\
NASA & National Aeronautics and Space Administration\\
\emph{N\!\!B} & \textit{Nota Bene}\\
NED & NASA/IPAC Extragalactic Database\\
NGC & New General Catalogue of Nebulae and Clusters of Stars\\
NYU & New York University\\
OH & Ohio\\
ORG & Organization\\
P & Page\\
PP & Pages\\
PADF & Pitch Angle Distribution Function\\
PASP & Publications of the Astronomical Society of the Pacific\\
PDF & Probability Density Function\\
PGC & Principal Galaxies Catalogue\\
PREP & Preparation\\
PUBL & Publications\\
RC3 & Third Reference Catalogue of Bright Galaxies\\
S$^4$G & Spitzer Survey of Stellar Structure in Galaxies\\
\textsc{sparcfire} & SPiral ARC FInder and REporter\\
SD & Standard Deviation\\
SMBH & Supermassive Black Hole\\
SOC & Society\\
SPIE & Society of Photo-Optical Instrumentation Engineers\\
UAE & United Arab Emirates\\
UCONN & University of Connecticut\\
ULX & Ultraluminous X-ray\\
US & United States\\
USA & United States of America\\
UV & Ultraviolet\\
VIC & Victoria\\
VOL & Volume\\
WWW & World Wide Web
\end{tabular}



}

\begin{adjustwidth}{-4cm}{0cm}
\printendnotes[custom] 

\reftitle{References}


\bibliography{bibliography.bib}

%


\end{adjustwidth}
\end{document}